\documentclass[11pt,a4paper]{article}
\usepackage{geometry}                
\usepackage{graphicx}
\usepackage{amssymb,amsmath,amsfonts}
\usepackage{amsthm}
\usepackage[dvipsnames]{xcolor}
\usepackage{bm}
\usepackage[colorinlistoftodos,prependcaption,textsize=small]{todonotes}
\usepackage{siunitx}
\usepackage[square,numbers]{natbib}

\usepackage{hyperref,color}
\usepackage{authblk}

\definecolor{webgreen}{rgb}{0,.35,0}
\definecolor{RoyalBlue}{rgb}{0,0,0.9}

\hypersetup{
	colorlinks=true, linktocpage=true, pdfstartpage=3, pdfstartview=FitV,
	breaklinks=true, pdfpagemode=UseNone, pageanchor=true, pdfpagemode=UseOutlines,
	plainpages=false, bookmarksnumbered, bookmarksopen=true, bookmarksopenlevel=1,
	hypertexnames=true, pdfhighlight=/O,
	urlcolor=magenta, linkcolor=RoyalBlue, citecolor=webgreen,
	pdfkeywords={},
	pdfcreator={pdfLaTeX},
	pdfproducer={LaTeX with hyperref}
}
\usepackage[capitalise]{cleveref}

\newcommand{\yp}[1]{{\color{black}{#1}}}

\newcommand{\ve}{\varepsilon}
\newcommand{\pa}{\partial}

\newcommand{\wf}{\widetilde F}
\newcommand{\wu}{\widetilde U}

\title{The Dynamics of Vesicles Driven Into Closed Constrictions by Molecular Motors}
\author[1]{Youngmin Park\footnote{Corresponding author \href{mailto:ypark@brandeis.edu}{ypark@brandeis.edu}}}
\author[1]{Thomas G. Fai}
\affil[1]{Department of Mathematics, Brandeis University, Waltham, MA 02453}

\date{}                                           

\begin{document}

%

\maketitle

\begin{abstract}
	We study the dynamics of a model of membrane vesicle transport into dendritic spines, which are bulbous intracellular compartments in neurons driven by molecular motors. We reduce the lubrication model proposed in [Fai et al., Active elastohydrodynamics of vesicles in narrow, blind constrictions. Phys. Rev. Fluids, 2 (2017), 113601] to a fast-slow system, yielding an analytically and numerically tractable equation equivalent to the original model in the overdamped limit. The model's key parameters are the ratio of motors that prefer to push toward the head of the dendritic spine to the ratio of motors that prefer to push in the opposite direction. We perform a numerical bifurcation analysis in these parameters and find that steady-state vesicle velocities appear and disappear through several saddle-node bifurcations. This process allows us to identify the region of parameter space in which multiple stable velocities exist. We show by direct calculations that there can only be unidirectional motion for sufficiently close vesicle-to-spine diameter ratios. Our analysis predicts the critical vesicle-to-spine diameter ratio, at which there is a transition from unidirectional to bidirectional motion, consistent with experimental observations of vesicle trajectories in the literature.
	
\end{abstract}

\section{Introduction}

\yp{Pyramidal neurons, the most ubiquitous type of neurons in the mammalian neocortex, each feature tens of thousands of excitatory convergent synaptic inputs. Most incoming synaptic signals terminate on sub-micron bulbs known as dendritic spines \cite{nimchinsky2002structure}. Spines exhibit a significant degree of morphological plasticity \cite{kasai2010structural,holtmaat2009experience} with pathological spine formation implicated in disorders such as Autism spectrum disorder and Alzheimer's disease \cite{penzes2011dendritic}. Normal synaptic function, including the dynamic process of spine remodeling, requires intracellular transport for maintenance \cite{da2015positioning}. Micron-sized vesicles carrying surface proteins are squeezed through the submicron-sized neck, undergoing strong deformations before fusing with the spine head. Recent experiments have shown that movement is not always unidirectional (translocation), but includes no movement (corking), and rejection \cite{park2006plasticity,wang2008myosin}. The mechanisms underlying these directional changes are not well understood. 
	
	To understand this question in greater detail, we use two primary considerations. First, we explicitly include intracellular transport in a vesicle trafficking model to understand how motor forces affect vesicle movement. This idea is common and there are many theoretical studies in this direction. Some use Markov processes representing motor complexes to understand the distribution of motor complex velocities \cite{muller2008tug,kunwar2011mechanical} and mean first passage times to transport targets on dendritic morphologies \cite{bressloff2009directed,newby2009directed,newby2010random,newby2011asymptotic,bressloff2013metastability}. More detailed studies include individual motors as part of a larger complex or population, which generatively produces bidirectional motion despite the assumption of symmetry \cite{julicher1995cooperative,guerin2011motion,allard2019bidirectional,portet2019deciphering}. However, the effect of constrictions on \yp{cargo} dynamics has typically been held constant or neglected. Indeed, intracellular movement into closed spaces is the second fundamental assumption of this study. 
	
	In contrast to open constrictions, which feature an infinitely long tube with a small constricted region, closed constrictions feature a semi-infinite tube that is closed on one end. How constricted and closed spaces affect molecular motor dynamics is a question that has been less studied. Earlier studies of constricted motion often consider open ends and are of general interest with related problems appearing in manufactured elastic capsules \cite{dawson2015extreme,duncanson2015microfluidic}, hydrogels \cite{li2015universal}, the movement of living cells \cite{bagnall2015deformability,byun2013characterizing,gabriele2010simple}, and axonal transport \cite{zimmermann1996accumulation,koehnle1999slow,walker2019local}. However, \cite{fai2017active} found that constrictions enrich motor populations' dynamics, allowing motors to switch between multidirectional and unidirectional motion. This paper aims to thoroughly explore the dynamics of the model by classifying the bifurcations of the underlying ODE.}


The paper is organized as follows. In Section \ref{sec:lubrication}, we briefly review the derivation of the lubrication model and its nondimensionalization, and we briefly discuss two equivalent versions, one of which (the ``slow'' subsystem) is the model considered in \cite{fai2017active}, and other of which (the ``fast'' subsystem) is the model we explore in-depth in this paper. In Section \ref{sec:bifurcations}, we numerically establish the existence and robustness of multistability through a bifurcation analysis of the ``fast'' subsystem. We corroborate some numerical results in Section \ref{sec:existence}, where we analytically establish the existence and stability of particular velocities as a function of key parameters. We conclude the paper with a discussion that includes estimates of realistic parameter regimes of this model, and the resulting behaviors predicted for different dendritic spines. All code, data, and documentation for reproducing figures in this paper are available on our GitHub repository at \url{https://github.com/youngmp/park_fai_2020}.

\section{Lubrication Model}\label{sec:lubrication}

\begin{figure}[ht!]
	\makebox[\textwidth][c]{
		\centering
		\includegraphics[width=.75\textwidth]{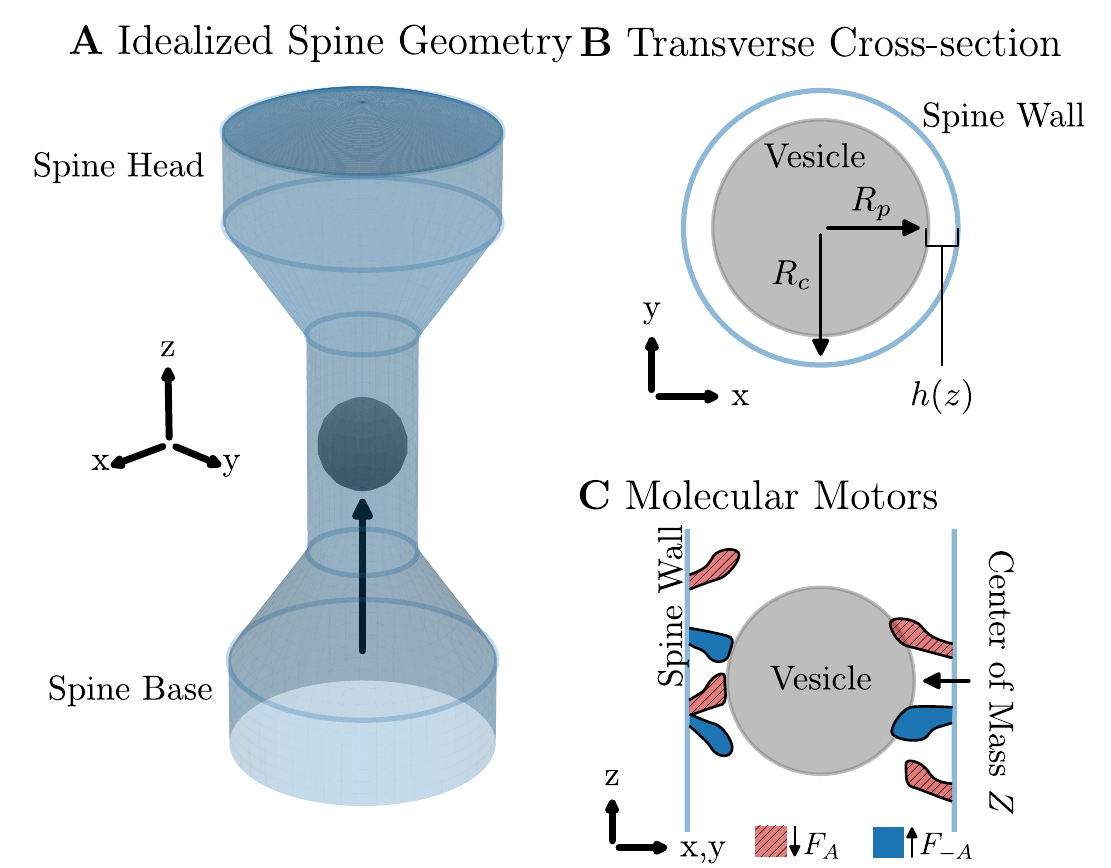}}
	\caption{\yp{Idealized dendritic spine and molecular motors. A: Three dimensional spine geometry with the vesicle (black sphere) shown at the center of the constriction. The black arrow shows the direction towards the spine head $R_p=0.96$, $R_c=1.22$. See Figure \ref{fig:constriction}A for additional details on the geometry. B: Transverse cross-section. C: Vertical cross-section with molecular motors. Blue: upwards-preferred motors. Red: downwards-preferred motors.}}\label{fig:geometry}
\end{figure}

\yp{The lubrication model we consider in this paper is an idealized model of vesicle translation through the neck of a dendritic spine}. In contrast to previous studies of transport through constrictions in periodic or unbounded tubes, we consider constrictions closed at one end to model transport into spine-like geometries \yp{(Figure \ref{fig:geometry}A)}.

\begin{figure}[ht!]
	\makebox[\textwidth][c]{
		\centering
		\includegraphics[width=1.25\textwidth]{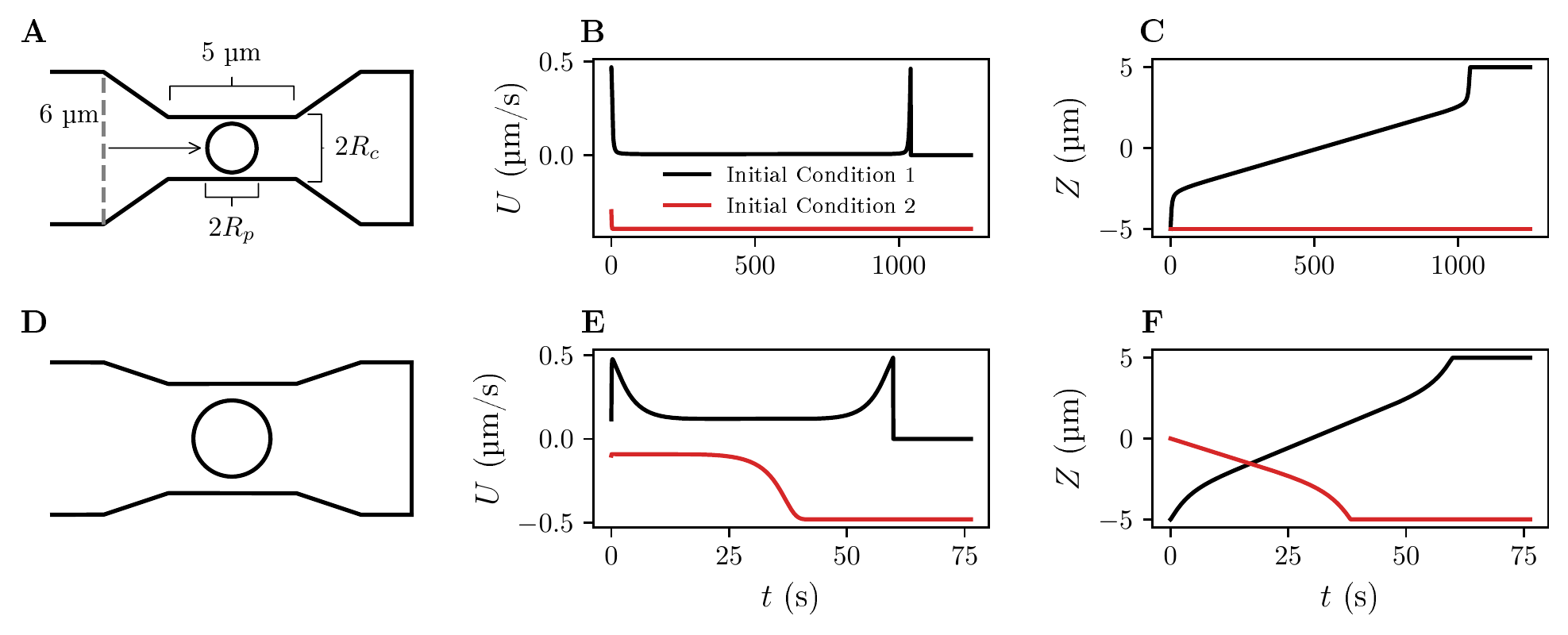}}
	\caption{Constriction geometry and resulting dynamics for different parameter sets. A: The initial spine diameter (6\si{.\um}) decreases to the neck radius $R_c=1.22$\si{.\um}. The vesicle (black circle, radius $R_p=0.96$\si{.\um}) begins at the base of the channel (dashed vertical gray line) and moves in the direction of the arrow for initial condition 1. B, C: Resulting velocity $U$ (\si{.\um/s}) and position $Z$ (\si{.\um}) plotted over time (s) for two different initial conditions $(U_0,Z_0)=(0.43 \si{.\um/s},-5\si{.\um})$ (black) and $(-0.3 \si{.\um/s},-5\si{.\um})$ (red). We use the parameters $\yp{\phi}=0.57$, $\pi_1=1$, $\pi_3=1$, $\pi_4=4.7$, $\pi_5=0.1$, $\pi_6=10$, $F_0=50$. D, E, F: same information as A, B, C, but with parameters $R_c=2.15$\si{.\um}, $R_p=1.5$\si{.\um}, $\yp{\phi}=0.54$, $\pi_1=1$, $\pi_3=1$, $\pi_4=4.7$, $\pi_5=0.02$, $\pi_6=10$, $F_0=200$. Initial spine diameter is 6\si{.\um}. The two initial conditions are $(U_0,Z_0)=(0.17 \si{.\um/s},-5\si{.\um})$ (black) and $(-0.1 \si{.\um/s},0\si{.\um})$ (red). Simulation parameters $\ve=1$, \texttt{dt}=0.02, integrated \yp{numerically} (see Appendix \ref{a:integration}).}\label{fig:constriction}
\end{figure}

Before deriving the lubrication model, we first define variables and parameters. The vesicle center of mass is $Z$ (Figure \ref{fig:geometry}C), and its radius is $R_p$ \yp{(Figure \ref{fig:geometry}B)}. The function $p(z)$ is the pressure exerted on the vesicle at position $z\in I(Z)$, \yp{where $I(Z) = [Z-R_p,Z+R_p]$ and the center of mass position $Z=0$ corresponds to the thinnest portion of the constriction (see Figure \ref{fig:constriction}A, B, where the vesicles (black circles) are drawn at position $Z=0$). The function $h(z)$ is the distance between the vesicle surface and the spine wall at position $z$ (Figure \ref{fig:geometry}B).
	
	We now closely follow the derivation of the lubrication model in \cite{fai2017active}. To start, we assume that the constriction radius $R_c$ is close to the vesicle radius $R_p$. Therefore, the minimum distance between the vesicle surface and spine wall is very small, i.e., $\min_{z\in I(Z)}\{h(z)\} \ll R_c$. In this scenario, fluid backflow surrounds objects entering closed constrictions. This backflow introduces a large velocity gradient, making particular terms dominate in the Navier-Stokes equations, allowing a reduction of the equations using lubrication theory \cite{acheson1991elementary}. Applied to our problem, lubrication theory yields,
	\begin{equation*}
	u(z) = \frac{1}{2\mu} \frac{\pa p(z)}{\pa z} r(r-h(z)) + \frac{U}{h(z)} r,
	\end{equation*}
	where $u(z)$ is the fluid velocity at position $z$, $r \in[0,R_c-R_p]$ is the radial coordinate in the thin fluid layer, $h(z)$ is the maximum radial thickness of the fluid layer at position $z$, and $U = dZ/dt$ is the vesicle velocity in the $Z$ direction. We will often refer to $h(z)$ as height, not to be confused with the $Z$-position.
	
	By incompressibility, the flux $Q$ through the gap must be equal through each cross-section, so
	
	\begin{equation}\label{eq:flux}
	Q = 2\pi R_c \left( -\frac{h^3}{12\mu}\frac{\pa p}{\pa z} + \frac{1}{2} U h \right)   = \text{const.}
	\end{equation}
	Rewriting \eqref{eq:flux} in terms of $\pa p/\pa z$ and integrating yields,
	\begin{equation}\label{eq:pressure}
	\frac{p(z)-p_0}{6\mu} = U \int_{Z-R_p}^z \frac{1}{h^2(s)} ds - \frac{2Q}{2\pi R_c} \int_{Z-R_p}^z \frac{1}{h^3(s)}ds.
	\end{equation}
	Setting $z=Z+R_p$ in Equation \eqref{eq:pressure} determines the flow rate $Q$ in terms of the pressure drop $\Delta p:=p(Z+R_p)-p_0$, which is a function of the applied force $F$ by $\Delta p = F/(\pi R_p^2)$. This results in the equation,
	\begin{equation}\label{eq:Q}
	Q = 2\pi \yp{R_c} \left(U \int_{\yp{I(Z)}} \frac{1}{h^2(s)} ds - \frac{F(U)}{6\pi \yp{R_p}^2\mu}\right)\left/ \left(2 \int_{\yp{I(Z)}} \frac{1}{h^3(s)} ds\right)\right. ,\\
	\end{equation}
	where $F(U)$ is the net force from the molecular motors and is a key feature that contributes to the tug-of-war dynamics of the model. By conservation of mass, the fluid dragged forward by the vesicle balances the backflow $Q$:
	\begin{equation}\label{eq:Q2}
	Q = -\pi R_c^2 U,
	\end{equation}
	where $R_c$ is the radius of the constriction (Figure \ref{fig:geometry}B). To close the system of equations for $p(z)$, $h(z)$, $Q$, and $U$, we define the constitutive law relating height to pressure:
	\begin{equation}\label{eq:h}
	h(z) = \tilde R_c(z) - \sqrt{R_p^2 - (z-Z)^2} + C[p(z)-p_0],
	\end{equation}
	where we assume that the vesicle is approximately spherical, $C$ is the compliance of the vesicle, and $\tilde{R}_c(z)$ is the radius of the channel at position $z$. Note that $\min_z\{\tilde{R}_c(z)\}=R_c$ as long as $0 \in I(Z)$. Equations \eqref{eq:flux}--\eqref{eq:h} constitute the reduced axisymmetric model of vesicle trafficking.
}


While the viscosity of water is on the order of $\mu=0.69$\si{.m.Pa.s} at a body temperature of 37\si{.\degreeCelsius}, proteins, filaments, and organelles densely pack the intracellular environment \cite{park2006plasticity,yuste2010dendritic,gray1959axo}, which may increase the effective viscosity by several orders of magnitude. To reflect this assumption, we take $\mu=1.2$\si{.m.Pa.s} \cite{fai2017active}.  The vesicles we consider in this paper are recycling endosomes, which serve to replenish surface proteins and vary from \yp{1--2\si{.\um}} in diameter \cite{da2015positioning}. Recycling endosomes are much larger than other vesicles commonly found in dendritic spines. Other vesicles are on the order of ten to hundreds of nanometers and may serve other functions in spines (based on data from VAST Lite \cite{berger2018vast}). For simplicity, and because we do not consider non-recycling endosomes, we will refer to recycling endosomes as vesicles throughout this paper.

\yp{Many studies assume constant forces \cite{adrian2014barriers,kusters2014forced}, but in reality the forces generated by molecular motors are dependent on quantities such as the cargo velocity. To capture this effect, we include a biophysical model of the forces generated by two species of myosin motors that are likely to dominate transport into spines \cite{da2015positioning}.} In our model, the two species are identical except that one prefers to push the vesicle up towards the spine head and the other prefers to push the vesicle down away from the spine head. 

\begin{figure}
	\makebox[\textwidth][c]{
		\centering
		\includegraphics[width=1.25\textwidth]{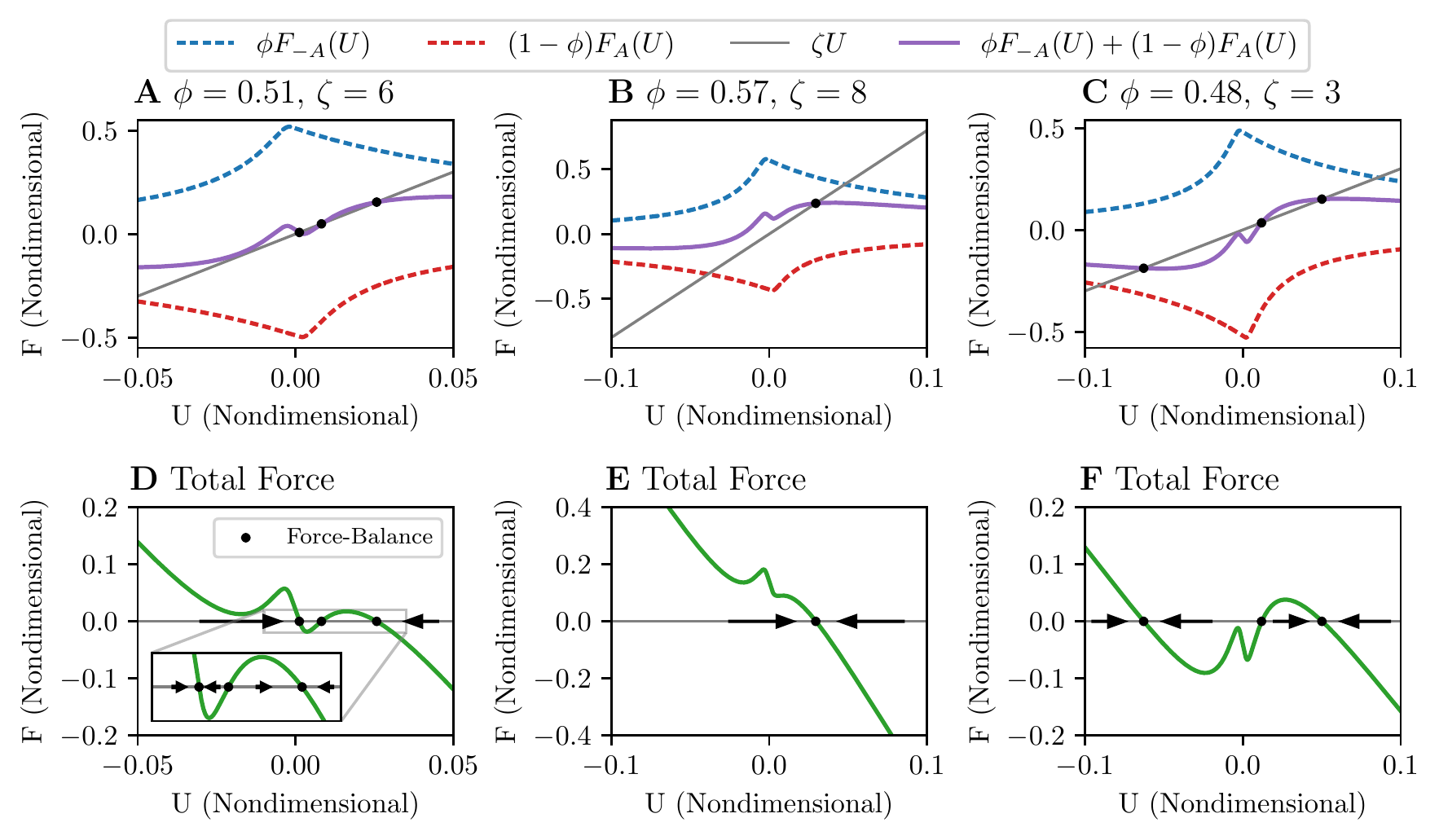}}
	\caption{\yp{Example force-velocity curves. A, B, C: as parameters $\phi$ and $\zeta$ vary, the underlying force-velocity curves and viscous drag forces change. Blue dashed: proportion of the motor force from upwards-pushing motors. Red dashed: proportion of the motor force from downwards-pushing motors. Purple: total force from molecular motors. Gray: total force from viscous drag. Black dots indicate intersections between the motor forces and viscous drag. D, E, F: respective total force including viscous drag, i.e., plots of $\phi F_{-A}(U)+(1-\phi)F_A(U) - \zeta U$. Black dots indicate force-balance (equilibria), and arrows indicate stability. The changing numbers of equilibria as a function of parameters indicate the loss or gain of multistability. We use the parameters $\pi_1=1$, $\pi_3=1$,$\pi_4=4.7$,$\pi_5=0.1$,$\pi_6=10$}}\label{fig:fv}
\end{figure}

\yp{Forces exerted by each species are described entirely using the standard convention of force-velocity curves, where the motor forces depend on the cargo velocity (Figure \ref{fig:fv})}. Following the notation of \cite{fai2017active}, the net motor force is written $F(U) = \phi F_{-A}(U) + (1-\phi) F_{A}(U)$, where $F_{-A}(U)$ and $F_A(U)$ are the force-velocity curves of motors that push towards and away from the spine head, respectively. The parameter $\phi$ represents the ratio of motor populations: $\phi=0$  corresponds to only downwards-pushing motors, $\phi=1$ corresponds to only upwards-pushing motors, and $\phi=0.5$ corresponds to equal numbers of motors pushing up and down.

For a given species, when the vesicle moves in the preferred direction, the motors attach and detach with intrinsic rates $\alpha$ and $\beta$, respectively. In the non-preferred direction, the motors not only detach due to the rate $\beta$ but are subject to yield effects: the motors extend up to a finite extension, beyond which the motors yield and no longer exert a force. The force $p(z)$ exerted by each motor depends on their position $z$ and is generally a monotonically increasing function with $p(0)=0$. In the present study, we use
\begin{equation}\label{eq:motor_force}
p(z)=p_1(e^{\gamma z}-1),
\end{equation}
\begin{figure}[ht!]
	\centering
	\includegraphics[width=.75\textwidth]{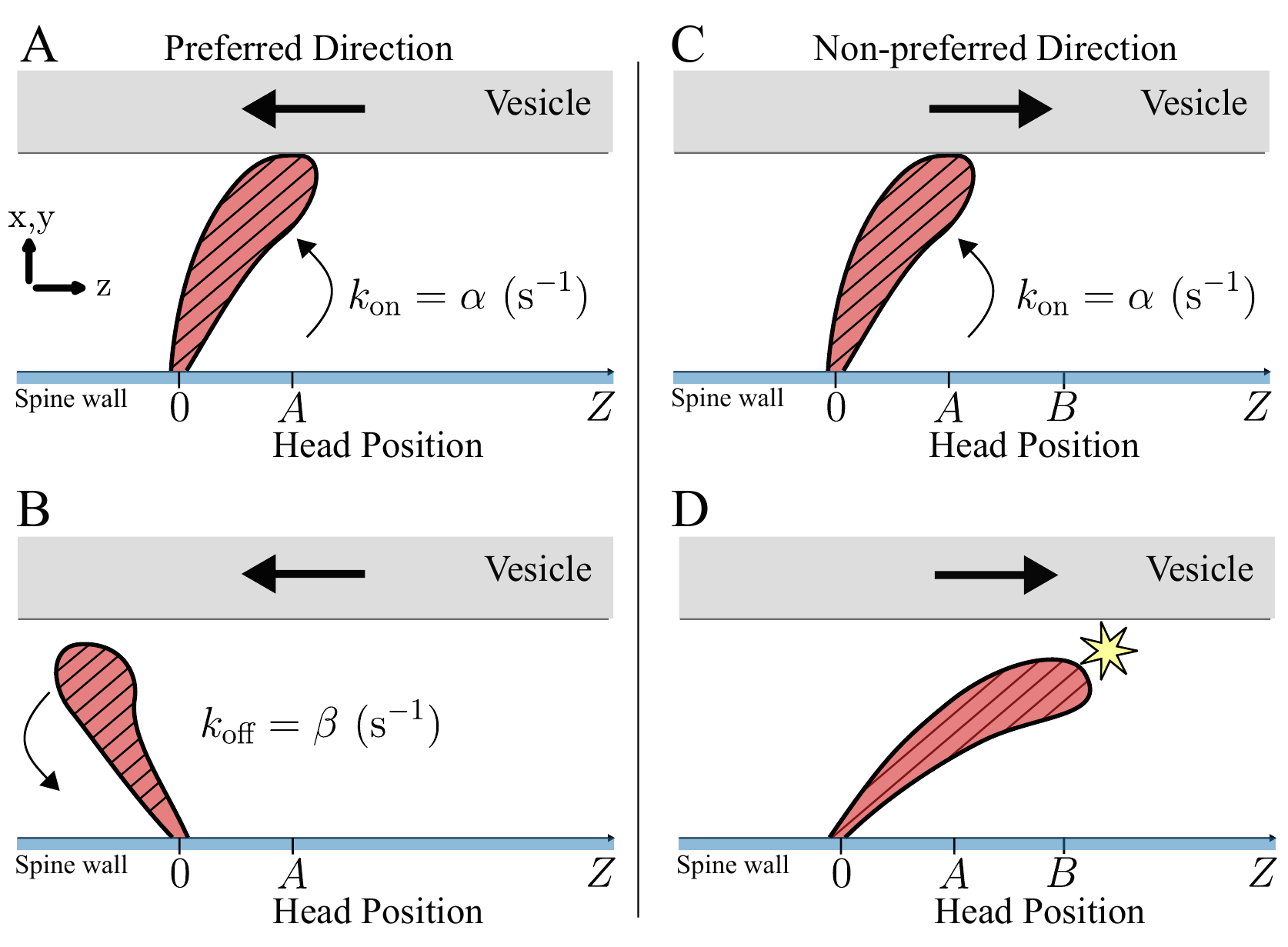}
	\caption{\yp{Microscopic motor dynamics of a downwards-preferred motor. The $x$-axis represents the $Z$ coordinate of the motor head relative to its base. A, B: when the vesicle (gray) moves in the preferred direction of a motor (red hatched), the motor attaches with a rate $\alpha$ and detaches with a rate $\beta$. C, D: when the vesicle moves in the non-preferred direction of a motor, the vesicle attaches with a rate $\alpha$ and detaches with rate $\beta$, but has an additional mechanism of detachment when the motor extends past $Z=B$}.}\label{fig:myosin}
\end{figure}
where $p_1$ and $\gamma$ are the motor force parameters (note that the position $z$ in Equation \eqref{eq:motor_force} represents the relative position of an individual motor, which is distinct from the $z$ used in the height function Equation \eqref{eq:h}. Figure \ref{fig:myosin} contains a brief description of the microscopic motor dynamics. Because we only focus on the mean-field dynamics, we will no longer reference $p(z)$, and any further reference to position $z$ will refer to the absolute position used in Equation \eqref{eq:h}). With this choice of force-extension, in the limit of large motor number, the forces in the preferred and non-preferred directions are functions of velocity. For upwards-pushing motors, the force-velocity curve, $F_{-A}(U)$, is given by
\begin{equation}
F_{-A}(U) = \begin{cases}
\frac{\alpha n_0 p_1}{\alpha c(U) + \beta} \frac{e^{\gamma A} \left(1-e^{-\beta(B-A)/U} e^{\gamma(B-A)}\right) - \left(1- e^{-\beta(B-A)/U}\right)(1+\gamma U/\beta)}{1+
	\gamma U/\beta}, & U < 0\\
\frac{\alpha n_0 p_1}{\alpha+\beta}\frac{e^{\gamma A}-1 - \gamma U/\beta}{1+\gamma U/\beta}, &  U \geq 0
\end{cases},\label{eq:fv0}
\end{equation}
where $c(U)=1-\exp[\beta(B-A)/U]$. Because the downwards-pushing motors follow the same rules, but for opposite signs in force and velocity, it follows that $F_{A}(U) = -F_{-A}(-U)$. We refer the reader to \cite{fai2017active,hoppensteadt2012modeling} for details on the derivation of the force-velocity functions $F_{-A}(U)$ and $F_{A}(U)$. \yp{We show the nondimensional version of the force-velocity curves (Equation \eqref{eq:fv}) in Figure \ref{fig:fv}}.

\subsection{Nondimensionalized Lubrication Model}

To enable an analysis of the dynamics of \cref{eq:pressure,eq:Q,eq:Q2,eq:h}, we first reduce the equations: we plug in Equation \eqref{eq:pressure} into Equation \eqref{eq:h} and Equation \eqref{eq:Q2} into Equation \eqref{eq:Q}, yielding a system of two equations for the velocity $U$ and the height between the vesicle and the constriction wall $h(z)$:
\begin{align}
U &= \frac{F(U)}{6\pi \yp{R_p} \mu}\frac{1}{\int_{Z-R_p}^{Z+R_p} \frac{\yp{R_p R_c}}{h^3(s)} + \frac{\yp{R_p}}{h^2(s)} ds},\label{eq:reduced_lub1}\\
h(z) &= \tilde R_c(z) - \sqrt{R_p^2 - (z-Z)^2} + C 6\mu \left[ U \int_{Z-R_p}^{z}\frac{1}{h^2(s)}ds - \frac{2Q}{2\pi R_c}\int_{Z-R_p}^{z}\frac{1}{h^3(s)}ds \right].\label{eq:reduced_lub2}
\end{align}
Next, we nondimensionalize Equations \eqref{eq:reduced_lub1} and \eqref{eq:reduced_lub2} and take $\widetilde z = z/R_p$, $\widetilde h = h/R_p$, and $\widetilde U = 6\pi R_p \mu U/F_0$, where $F_0 = (\exp(\gamma A)-1)\alpha p_1 n_0 / (\alpha + \beta)$ is the stall force. Note that here we use tildes to denote dimensionless quantities (tildes will be dropped later on). Plugging the nondimensionalized terms into Equations \eqref{eq:reduced_lub1} and \eqref{eq:reduced_lub2} yields,
\begin{align}
\widetilde U &=\wf(\wu)/\tilde \zeta(\widetilde Z) \label{eq:nondim_u},\\
\widetilde h(\widetilde Z + \widetilde z) &= \tilde R_c(\widetilde Z+    \widetilde z)/R_p - \sqrt{1-\widetilde{z}^2} + \pi_2 \widetilde U \int_{-1}^{\widetilde{z}} \left[\widetilde h^{-2}(\widetilde Z + s) + \yp{\pi_1}\widetilde h^{-3}(\widetilde Z + s) \right]ds\label{eq:nondim_h},
\end{align}
where \yp{$\pi_1=R_c/R_p$}, $\pi_2 = CF_0/(\pi R_p^3)$, and
\begin{equation}\label{eq:zeta}
\tilde \zeta(\widetilde Z) = \int_{-1}^1 \left[\widetilde h^{-2}(\widetilde Z + s) + \yp{\pi_1} \widetilde h^{-3}(\widetilde Z + s) \right]ds.
\end{equation}
The function $\tilde \zeta$ is the viscous drag coefficient produced by the constriction geometry. We show examples of this function and discuss its importance for this system's dynamics in the next section. \yp{Although the nondimensional drag term $\tilde{\zeta}$ (Equation \eqref{eq:zeta}) is purely geometrical, the drag force itself is a direct consequence of the fluid flow. The dimensional drag in Equation 2.8 includes a factor of $6\pi R_p \mu$, and the fluid viscosity $\mu$ has been absorbed into the nondimensionalization. The purely geometrical term $\tilde{\zeta}$ is simply a correction factor to the Stokes drag law that quantifies the degree of confinement. Indeed, this is a consequence of the higher force needed to sustain the large velocity gradients in the narrow gaps.}  

The nondimensionalized net motor force $\wf(\wu) = \phi \wf_{-A}(\wu) + (1-\phi)\wf_A(\wu)$ consists of two functions $\wf_{-A}(\wu)$ and $\wf_A(\wu)$, which are related by $\wf_A(\wu)=-\wf_{-A}(-U)$ using the same symmetry arguments in the dimensional equations. It is straightforward to show that the nondimensionalized force-velocity curve is
\begin{equation}\label{eq:fv} 
\widetilde F_A (\wu) = \left\{ \begin{matrix}
-\frac{1+\pi_6\widetilde U(e^{\pi_4}-1)^{-1}}{1-\pi_6\widetilde U}, & \text{if}\quad \widetilde U < 0\\
\frac{-(\pi_3+1)}{\pi_3(1-e^{-\pi_5/\pi_6 \widetilde U}) + 1} \frac{[e^{\pi_4}(1-e^{\pi_5}e^{-\pi_5/\pi_6\widetilde U})]-(1-\pi_6\widetilde U)(1- e^{-\pi_5/\pi_6 \widetilde U})}{(e^{\pi_4}-1)(1-\pi_6\widetilde U)}, & \text{if} \quad \widetilde U \geq 0.
\end{matrix}
\right.
\end{equation}
The function includes numerous parameters representing various microscopic motor properties: $\pi_3 = \alpha/\beta$ (ratio of attachment and detachment rates), $\pi_4 = \gamma A$ (the nondimensional attachment position), $\pi_5 = \gamma(B-A)$ (the maximum displacement of a motor in its non-preferred direction), and $\pi_6 = (F_0/[6\pi R_p \mu])/(\beta/\gamma)$ (the ratio of velocity scales between translocation and motor adhesion dynamics). When conversions back to dimensional forces are needed, we write $F_X = \wf_X F_0$ for $X=A,-A$. \yp{For a fixed set of $\pi_i$, $i=3,4,5,6$, we show how the force-velocity curves change as a function of $\phi$ and $\zeta$}.

\subsection{Fast-Slow Lubrication Model}

From this point on, we work exclusively with the nondimensional system unless explicitly stated. Therefore, we write $Z = \widetilde Z$, $U = \wu$, $h=\widetilde h$, $\zeta=\tilde\zeta$ and $F=\wf$. Note that Equation \eqref{eq:nondim_h} includes a term to account for vesicle compliance with a prefactor of $\pi_2$. Representative parameters of vesicle compliance \yp{in the case of a spherical vesicle} reveal the nondimensional compliance $\pi_2$ to be relatively small, on the order of $\pi_2 \approx 0.09$ \cite{fai2017active}. To a first approximation, we take $\pi_2 \approx0$, which significantly simplifies Equations \eqref{eq:nondim_u} and \eqref{eq:nondim_h}. \yp{This approximation means that we assume a rigid, spherical vesicle. We refer the reader to the discussion in Section \ref{sec:limitations} for details regarding this choice}.

The low compliance limit yields the fast-slow system,
\begin{equation}\label{eq:fs1}
\begin{split}
\frac{dZ}{dt} &= U,\\
\ve\frac{dU}{dt} &= F(U) - \zeta(Z) U.
\end{split}
\end{equation}
$F$ is the dimensionless net motor force, $U$ is the dimensionless vesicle velocity, $Z$ is the dimensionless vesicle position, $\zeta$ is the dimensionless drag that captures information about the constriction geometry (Equation \eqref{eq:zeta}), and $\ve$ is a dimensionless mass term that may be zero, and equals zero in the overdamped limit.

Figure \ref{fig:constriction} shows some example dynamics of Equation \eqref{eq:fs1}. Figure \ref{fig:constriction}A shows the axisymmetric idealized dendritic spine from Figure \ref{fig:geometry}. The base of the spine is marked by a vertical gray dashed line positioned at the dimensional position of $-5$\si{.\um}, with a base diameter of $6$\si{.\um}. The spine transitions linearly into the constriction, which has a radius of $R_c=1.22$\si{.\um}, and a length of  $5$\si{.\um}. The vesicle, shown as a black circle, has a radius of $R_p=0.96$\si{.\um}. The first initial condition we consider starts the vesicle at the base of the spine with positive initial velocity $(U_0,Z_0)=(0.43 \si{.\um/s},-5\si{.\um})$. Black curves show solutions to this initial condition. As the vesicle moves into the constriction, confinement effects at the neck significantly reduce the translocation velocity (Figure \ref{fig:constriction}B, black). However, the vesicle position increases until it reaches the end of the channel (Figure \ref{fig:constriction}C, black). We show another initial condition that starts at the base with a negative initial velocity $(U_0,Z_0)=(-0.3 \si{.\um/s},-5\si{.\um})$ and denotes solutions of this initial condition in red. The vesicle velocity remains negative until it hits the no-penetration boundary condition, which we impose at the two ends of the channel. When the vesicle hits the base of the spine at $-5$\si{.\um}, the velocity is instantaneously reset to zero and the position to $-5$\si{.\um}. This zero-velocity solution is in the basin of attraction of a negative velocity, so the vesicle remains at the base. Different initial conditions reveal different long-time dynamics, suggesting multistability. 

We show another representative spine in Figure \ref{fig:constriction}D, where the vesicle and constriction values are 1.5\si{.\um} and 2.15\si{.\um}, respectively (all other geometry parameters are the same as in Panel A). The two initial conditions are  $(U_0,Z_0)=(0.17 \si{.\um/s},-5\si{.\um})$ (black) and $(-0.1 \si{.\um/s},0\si{.\um})$ (red). Note that while the first initial condition (black) successfully translocates, the second initial condition (red) does not (Figure \ref{fig:constriction}F). Again, the differing dynamics as a function of initial conditions suggest the system is multistable. We remark that while we use piecewise linear channels throughout this paper, any constriction geometry is allowed as long as the vesicle radius is close to the channel radius, i.e., $R_p\approx R_c$.

As a starting point for our analysis, we explore system \eqref{eq:fs1} under two equivalent limits that reveal different aspects of the dynamics. Viewed in the ``slow'' time $t$, taking the \yp{overdamped} limit $\ve \rightarrow 0$ yields the slow subsystem, 
\begin{equation}\label{eq:slow}
\begin{split}
\frac{dZ}{dt} &= U,\\
0 &= F(U) - \zeta(Z) U.
\end{split}
\end{equation}
The dynamics of Equation \eqref{eq:slow} exist on the critical manifold $S_0$ defined by
\begin{equation}\label{eq:s0}
S_0 := \{(Z,U) \in \mathbb{R}^2 \,|\, 0 = F(U) - \zeta(Z)U\}.
\end{equation}
\yp{Equations \eqref{eq:slow} and \eqref{eq:s0} correspond to how the vesicle moves in real-time. Time $t$ is in units of seconds, and when re-dimensionalized, the center of mass of the vesicle moves with velocity $dZ/dt$ in \si{\um/s}. Translocation across a spine tends to occur on the order of minutes. On the other hand, individual myosin motors attach and detach on a time scale in the range of \si{10^{-1}.s} to \si{10^{-2}.s}. With a few dozen myosin motors operating at this time scale within a relatively viscous environment, force balance is virtually instantaneous relative to translocation. Thus we take the forces to satisfy the instantaneous force-balance condition in Equation \eqref{eq:s0}.}

\yp{This paper aims to classify the bifurcations of the manifold satisfying Equation \eqref{eq:s0} to understand how velocities become multistable}. Fenichel theory guarantees that for $\ve$ sufficiently small, the dynamics of Equation \eqref{eq:fs1} closely follow the dynamics on the slow manifold \cite{fenichel1979geometric,broer2013geometric}. While in the present study, we operate primarily within the overdamped limit where $\ve=0$, we sometimes take $\ve>0$ to numerically integrate the equations using standard methods, as in Figure \ref{fig:constriction}. We refer the reader to Appendix \ref{a:integration} for details of this approach.
\begin{figure}[ht!]
	\centering
	\makebox[\textwidth][c]{
		\includegraphics[width=1.2\textwidth]{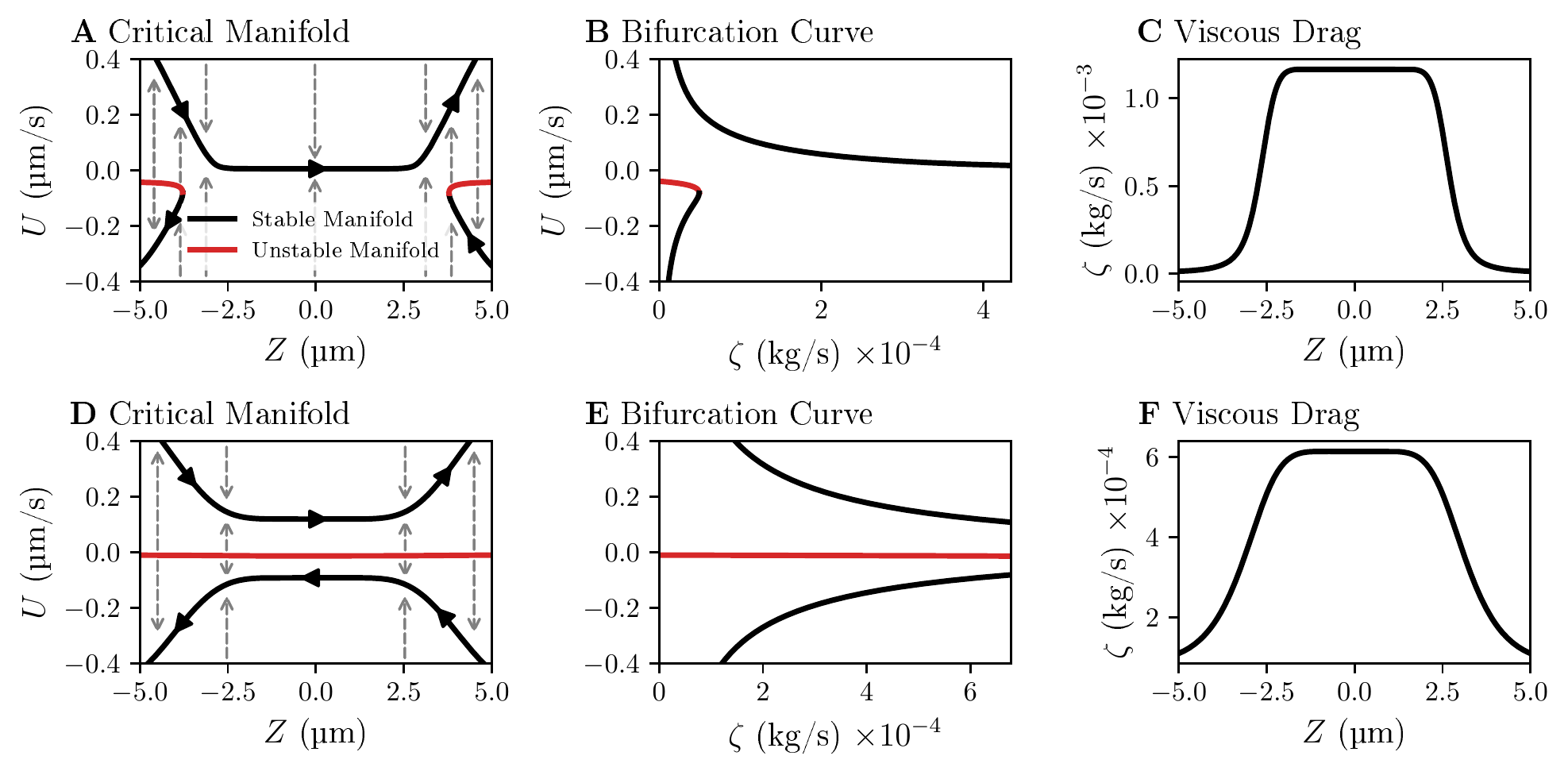}}
	\caption{The mapping between the bifurcation diagram and critical manifold through the viscous drag function. A: An example of the critical manifold in the phase space of $U$ and $Z$. Black arrowheads denote the direction of motion on the slow manifold. Gray dashed arrows indicate the direction of motion in the fast system. B: An example of a one-parameter bifurcation diagram. Steady-states $U$ are plotted as a function of $\zeta$. C: The relationship between viscous drag $\zeta$ and position $Z$. The dimensional positions $Z=-5$\si{.\um} through $Z=0$\si{.\um}, we expect the critical manifold to resemble a version of the bifurcation diagram given by the mapping between drag $\zeta$ and position $Z$. Beyond the constriction, from dimensional positions $Z=0$\si{.\um} through $Z=5$\si{.\um}, the critical manifold resembles a reflected version of the bifurcation curve. A--C: Parameters as in Figure \ref{fig:constriction}A--C. D--F: Parameters as in Figure \ref{fig:constriction}D--F.}\label{fig:c0}
\end{figure}
In Figure \ref{fig:c0}A,D, we show examples of the critical manifold for the corresponding geometries in Figure \ref{fig:constriction}A,D. \yp{We  compute the critical manifolds using the following process. We define a grid in $U$ and $Z$ and plot the value of the function $G(U,Z):=F(U) - \zeta(Z)U$. We then have a surface plot of the function $G(U,Z)$ on the $U$-$Z$ domain, from which we extract the contours where $G(U,Z)=0$. These contours give us the critical manifolds because they correspond precisely to the critical manifold condition $0 = F(U) - \zeta(Z)U$ in Equation \eqref{eq:s0}}.

A hybrid system determines the dynamics on the manifold: for a given set of initial conditions $(Z_0,U_0)$, the fast dynamics instantaneously carry the solution to the nearest stable manifold. Along the stable manifold, the slow dynamics evolve according to \eqref{eq:slow}, until the solution reaches a fold, at which point the fast dynamics instantaneously carry the solution to the next stable manifold. Figure \ref{fig:c0}A,D show these hybrid dynamics for the fast dynamics (dashed gray arrows) and for the slow dynamics (black arrows).

\yp{Let $s=t/\ve$, and call $s$ the ``fast'' time. A straightforward application of the chain rule yields,
	\begin{align}
	\frac{dZ}{ds} &= \ve U\\
	\frac{dU}{ds} &= F(U) - \zeta(Z) U.
	\end{align}
	Thus, from the perspective of the fast time $s$, $Z$ is a slow variable. Letting $\ve \rightarrow 0$ yields the fast subsystem,
	\begin{align*}
	\frac{dZ}{ds} &= 0\\
	\frac{dU}{ds} &= F(U) - \zeta(Z) U.
	\end{align*}
	The fast subsystem describes the same overdamped limit as the slow subsystem, but instead of viewing the vesicle position in real-time and the velocity dynamics as instantaneous, it freezes the vesicle position and shows how the velocity dynamics converge to force-balance in the timescale of molecular motors. Unlike the slow subsystem, the fast subsystem directly informs us of the stability of equilibria. Moreover, the fast subsystem yields a substantially more tractable version of the lubrication model that is much easier to analyze using bifurcation theory. Note that because $Z$ is constant in this limit and because $\zeta$ only depends on $Z$, what remains is the one-dimensional ODE,
	\begin{equation}\label{eq:fast}
	\frac{dU}{ds} = F(U) - \zeta U,
	\end{equation}
	where $\zeta$ can be treated as a parameter.}

The bifurcations of Equation \eqref{eq:fast} are related to the slow subsystem (Equation \eqref{eq:slow}) through the viscous drag term. Because $\zeta$ is a function of position $Z$, there exists a mapping from the critical manifold to the bifurcation curve. For example, as the vesicle center of mass approaches the center of the constriction, viscous drag increases monotonically (Figure \ref{fig:c0}C). At this stage, the bifurcation curve and critical manifold closely resemble scaled versions of each other (Figures \ref{fig:c0}A, B and Figure \ref{fig:c0}D, E when $Z\in[-5\si{.\um},0\si{.\um}]$). Beyond the center of the constriction, the viscous drag term decreases monotonically, and the critical manifold resembles a reflected version of the bifurcation curve (Figure \ref{fig:c0}A, B and Figures \ref{fig:c0}D, E when $Z\in[0\si{.\um},5\si{.\um}]$). Thus understanding the bifurcation curves and the viscous drag terms are sufficient to understand the critical manifold of the overdamped system. With this mapping in mind, we turn to a thorough numerical analysis of this system's bifurcations.

\section{Bifurcations of the Force-Velocity Curve}\label{sec:bifurcations}

\begin{figure}[ht!]
	\centering
	\includegraphics[width=\textwidth]{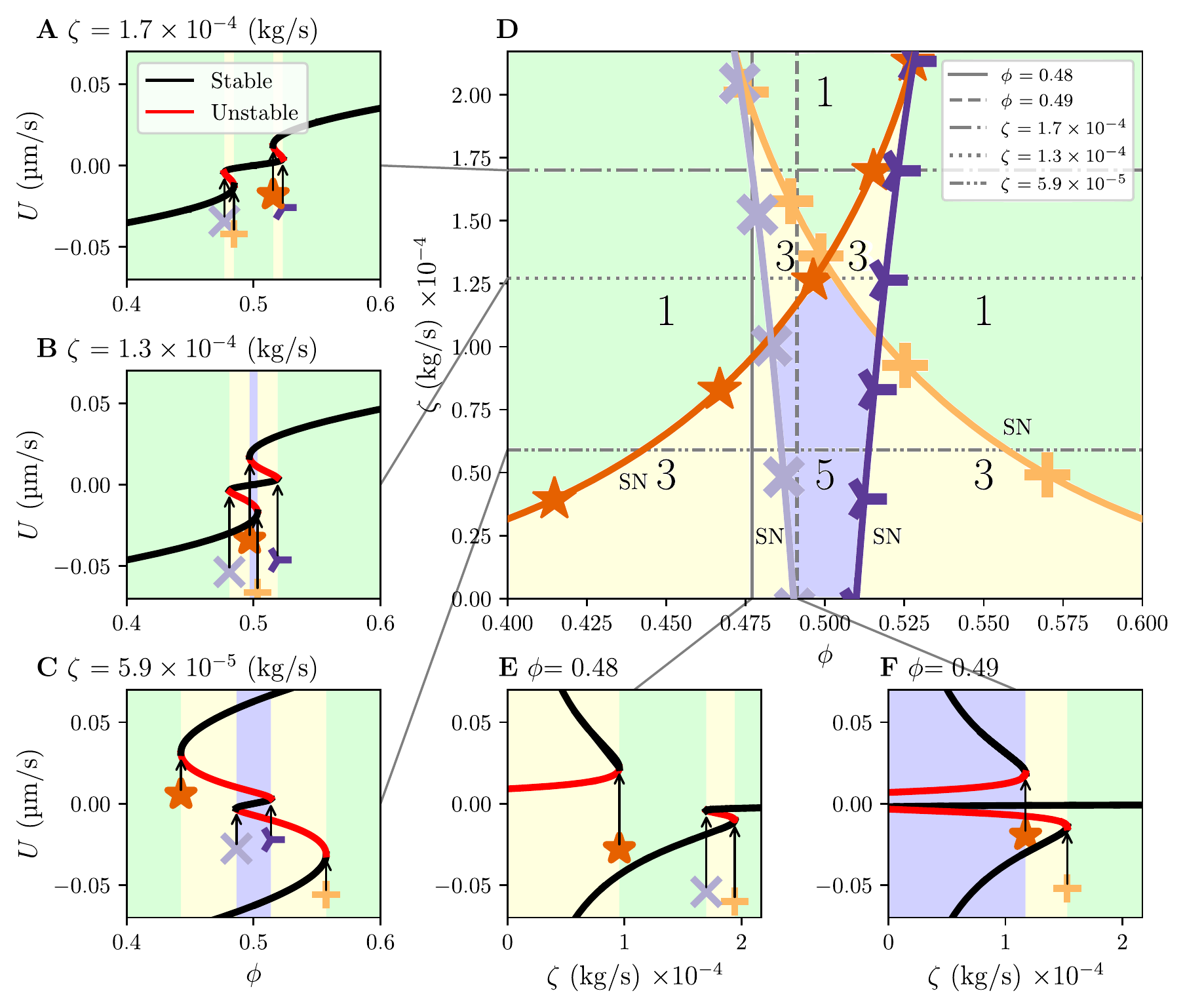}
	\caption{Two parameter bifurcation diagram in $\phi$ and $\zeta$. Saddle-node (SN) bifurcations are shown in (D) as colored branches with a unique color and symbol. Numbers in (D) indicate the total number of fixed points in the corresponding region of parameter space. Subplots A, B, C, E, F, show one-parameter slices of the two-parameter diagram. Saddle-nodes are labeled with the corresponding branch color and symbol. The critical vesicle-to-spine diameter ratio at the cusps is roughly 2\si{.\um}/3\si{.\um}.}\label{fig:2par}
\end{figure}
In this section, we perform a numerical bifurcation analysis of the fast subsystem by following the roots of the right-hand side of Equation \eqref{eq:fast}:
\begin{equation}
f(U) = \phi F_{-A}(U) +(1-\phi)F_{A}( U) - \zeta U.
\end{equation}
Details of the numerics are given in Appendix \ref{a:continuation}.

\subsection{Bifurcations in $\phi$-$\zeta$}

We begin with single-parameter bifurcation diagrams in $\phi$ and $\zeta$ by fixing one parameter and varying the other. In Figures \ref{fig:2par}A--C, we fix $\zeta$ at three different values and follow equilibria as a function of \yp{$\phi$}. The symmetry of these curves about $\phi=0.5$ comes from our choice of force-velocity curves for competing motors: we use identical force-velocity curves for both species, so the existence of fixed points for $\phi>0.5$ is the same but with opposite sign when $\phi<0.5$.

As $\zeta$ decreases from Panels A to C, the saddle-node denoted by the orange star occurs at progressively smaller values of \yp{$\phi$}, and the saddle-node denoted by the yellow $+$ occurs at progressively greater values of \yp{$\phi$}. The change results in the creation of multistable velocities (stable velocities are black and unstable velocities are red). In Panel A, there exist values of $\phi$ with only one or three fixed points. In Panels B and C, the change in the folds' position gives way to the existence of five fixed points. We also find that $\phi$ above and below 0.5 tend to yield positive and negative velocities. There are some exceptions, such as Panel C, where $\phi>0.5$ can result in negative velocity. This observation is the well-established tug-of-war effect. These one-parameter bifurcation diagrams give us a good starting point of how stable velocities change as a function of two parameters and some insight into the shape of the bifurcation surface. Note that in Panel A, the parameter range in $\phi$ for which there is only one solution is much greater relative to the range of $\phi$ for three solutions. We call the single-velocity solutions more ``robust'' relative to the multistable solutions. We can also conclude from Panel C that the tug-of-war effect is more likely to be seen for lower viscous drag values. We will later discuss robustness in a similar way, where relatively larger parameter ranges correspond to increased robustness.

In Figures \ref{fig:2par}E, F, we fix \yp{$\phi$} at 0.48 and 0.49, respectively, and vary $\zeta$. These bifurcation diagrams are less intuitive but can be understood as slices through the bifurcation surface described above. More importantly, these bifurcation diagrams reveal some information about the underlying critical manifold. In \ref{fig:2par}E, we see that if the vesicle has a sufficiently positive initial velocity at low drag, it maintains a positive velocity as the vesicle moves into the spine neck, and the drag grows due to the constriction. At a critical drag denoted by the orange star, the vesicle instantaneously switches velocity in the opposite direction by jumping down to the lower stable branch and eventually exits the constriction through the base of the spine at $-5$\si{.\um}. Figure \ref{fig:2par}F exhibits similar discontinuous behavior: with an appropriate positive initial velocity, the vesicle moves towards the constriction before jumping down to the stable middle branch where the velocity is near zero. In this scenario, the vesicle remains stuck for long times. Note that by the symmetry of these functions, the bifurcation diagrams look identical with a change of sign for $\phi=0.52,0.5$. Therefore, the model predicts that motor-driven transport through constrictions will generally push the vesicle towards the spine head as long as the initial condition is sufficiently far into the constriction and that upwards-pushing motors are dominant.

While one-parameter diagrams are useful, we wish to understand how multistability changes in the entire $\phi$-$\zeta$ parameter space. We address the question of multistability by noting how each one-parameter bifurcation changes as a function of an additional parameter. This process naturally partitions the $\phi$-$\zeta$ parameter space into multistable regions. Noting that our one-dimensional system only produces saddle-node (SN) bifurcations that produce or destroy pairs of fixed points, we only need to track these saddle-node bifurcations. This process yields Figure \ref{fig:2par}D, in which we suppress fixed points and only display bifurcation points and the number of fixed points in each region. Each of the four colored curves corresponds to a saddle-node bifurcation with a unique color and symbol. As expected, the number of fixed points changes across the various saddle-node curves. For example, panel A shows that as \yp{$\phi$} increases, the total number of fixed points changes in the order $1\rightarrow 3 \rightarrow 1 \rightarrow 3 \rightarrow 1$. The same can be observed in panel A by tracing the slice at $\zeta = 1.75\times 10^{-4}$ \si{kg/s}.

The two-parameter diagram completely characterizes the total number of fixed points for each region of the parameter space and shows that we can expect multistability in much of the displayed parameter space. If viscous drag is sufficiently small (in the bottom region of panel D), multistability exists for a wide range of motor ratios \yp{$\phi$}. As viscous drag increases, the range of multistability becomes much smaller as fixed points disappear through saddle-node bifurcations. For sufficiently large viscous drag $\zeta$, multistability ceases to exist as the saddle-nodes disappear through cusps, and there exists only one stable velocity for all motor ratios. The critical vesicle-to-spine diameter ratio at the cusps is roughly 2\si{.\um}/3\si{.\um}.

In terms of the bifurcation surface, the parameter slices in panels A--C show that for several values of fixed $\zeta$, the bifurcation surface contains four folds. By choosing greater values of $\zeta$, we find that the folds of the surface eventually flatten out through a pair of cusp bifurcations. Beyond this cusp bifurcation, for greater values of $\zeta$, we expect that the velocities $U$ are negative when $\yp{\phi} < 0.5$, positive when $\yp{\phi}>0.5$, and zero when $\yp{\phi} = 0.5$. From this geometric intuition, it follows that when confinement effects are sufficiently large, the vesicle's velocity is determined purely by the ratio of upwards- and downwards-pushing motors. We make this observation more rigorous in Section \ref{sec:existence}.

\subsection{Robustness in $\pi_4$, $\pi_5$}
\begin{figure}[ht!]
	\centering
	\includegraphics[width=\textwidth]{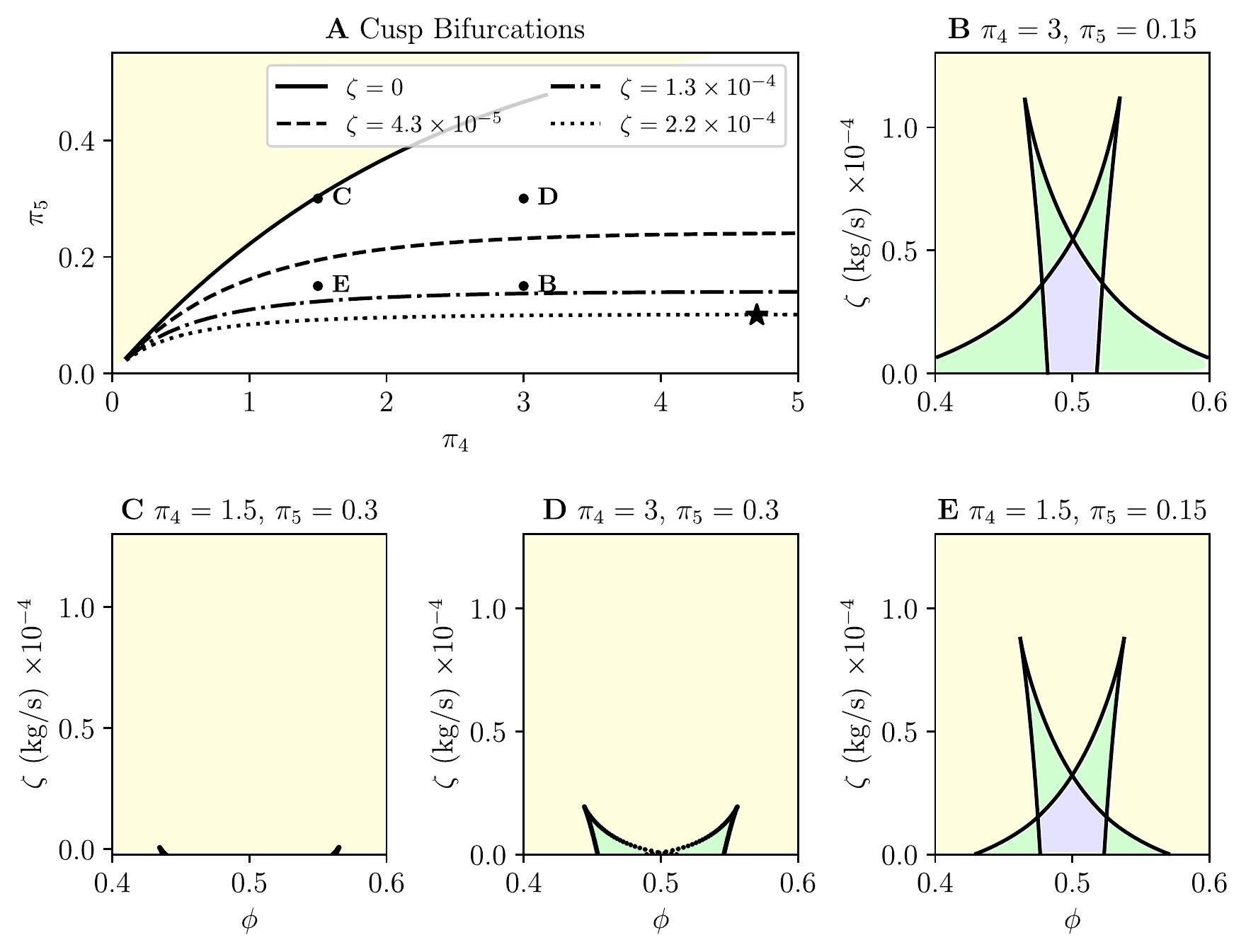}
	\caption{Cusp bifurcations as a function of $\pi_4$ and $\pi_5$. A: For each $\zeta>0$,cusp bifurcations exist along a set of $\pi_4,\pi_5$. Example level curves are plotted for $\zeta = 0,4.3\times10^{-5}, 1.3\times10^{-4},2.2\times10^{-4}$. We take 4 representative pairs of $\pi_4,\pi_5$ labeled B--E and show the corresponding two-parameter bifurcation diagrams in B--E. The point labeled with $\star$ corresponds to Figure \ref{fig:2par}.}\label{fig:cusps}
\end{figure}

When we fix the motor parameters $\pi_1$-$\pi_6$, Figure \ref{fig:2par} provides a complete description of how multistability changes as a function of motor ratio and constriction geometry. However, there may be variations in motor parameters due to the existence of multiple motor types such as myosin V and VI \cite{da2015positioning} and variations in ATP and ADP concentration, which is known to differentially modulate myosin motor dynamics \cite{zimmermann2015actin}.

As explained in detail in Appendix \ref{a:cusps}, the cusp bifurcation separates the parameter space between multistable velocities and globally stable or unstable solutions. Indeed, cusps generally serve as a sufficient condition for the existence of hysteresis. Therefore, understanding the cusp bifurcation may provide essential insights into controllability. For a given $\zeta$, it is possible to track cusps as a function of $\pi_4$ and $\pi_5$. The result of this process for various $\pi_4$ and $\zeta$ is shown in Figure \ref{fig:cusps}A. Each curve represents a cusp bifurcation as a function of $\pi_4,\pi_5$ for a given $\zeta$ (we briefly describe how we determined the location of these cusps in Appendix \ref{a:cusps}). 

Given parameters $\pi_4,\pi_5$ chosen somewhere in the $\pi_4$-$\pi_5$ parameter region below the $\zeta=0$ level curve in Figure \ref{fig:cusps}, the $\zeta$ level curves suggest that there exists a cusp bifurcation at some $\zeta^*(\pi_4,\pi_5)>0$. Because cusps are a sufficient condition for multistability, it follows that multistable states exist for appropriate choices of \yp{$\phi$} and $\zeta \leq \zeta^*$. The $\zeta=0$ level curve represents a sufficient condition for the loss of multistability: noting that for each fixed point, $\pa \zeta/\pa\pi_5 <0$ in a neighborhood about the fixed point, i.e., $\zeta$ decreases as $\pi_5$ increases, it follows that for any $\pi_4,\pi_5$ in at least a neighborhood above this curve, there can exist no cusp bifurcation with a positive $\zeta$. 

Various points in Figure \ref{fig:cusps}A are marked B--E, with corresponding two-parameter bifurcation diagrams shown in the remaining subplots, Figures \ref{fig:cusps}B--E. The point marked by $\star$ in Figure \ref{fig:cusps}A represents the $\pi_4$ and $\pi_5$ values for Figure \ref{fig:2par}. These diagrams show that the region of multistability tends to increase for smaller $\pi_5$ and greater $\pi_4$. Recalling that $\pi_5$ is the maximum displacement of a motor in its non-preferred direction, smaller $\pi_5$ implies that bidirectional motion due to noise can be made more likely by allowing the motor to detach earlier. Next, $\pi_4$ is the initial motor attachment position in either the preferred or non-preferred direction, and we find that the area of multistability increases as motors have a greater initial extension. Together, we predict that strong initial attachment forces combined with greater yield effects can result in more frequent directional switching.

\section{Existence and Stability of Solutions}\label{sec:existence}

\subsection{Existence}
The existence of a stationary solution $U = 0$ is straightforward to prove by inspection for $\phi=0.5$. In this section, we expand about this solution to determine the existence of solutions when $\phi$ is near the equal motor ratio $\phi=0.5$ and when $\zeta$ is large. We let $\phi = 0.5+\hat\phi$ and $\hat\zeta=1/\zeta$ and explore the cases where $\hat \phi$ and $\hat\zeta$ are small. These limits inform us of linear behavior of the velocity function $U=\yp{U}(\hat\phi,\hat\zeta)$ by writing
\begin{equation*}
\yp{U}(\hat\phi,\hat\zeta) = \left.\frac{\pa \yp{U}}{\pa \hat\phi}\right|_{(0,0)}\hat\phi + \left.\frac{\pa \yp{U}}{\pa \hat\zeta}\right|_{(0,0)}\hat\zeta + O(\hat\phi \hat\zeta,\hat\phi^2,\hat\zeta^2).
\end{equation*}

First we derive an equation for the small deviation $\phi=0.5+\hat \phi$, where $0<|\hat\phi|\ll1$. Constant velocity solutions $U=\yp{U}$ must satisfy
\begin{align*}
0 &= \phi F_{-A}(\yp{U}) + (1-\phi) F_{A}(\yp{U}) - \zeta \yp{U}\\
&= \frac{1}{2} F_{-A}(\yp{U}) + \frac{1}{2}F_A(\yp{U}) + \hat \phi [F_{-A}(\yp{U}) - F_A(\yp{U})] - \zeta \yp{U}.
\end{align*}
Solving for $\hat \phi$ yields,
\begin{equation*}
\hat \phi = \frac{1}{2}\frac{F_{-A}(\yp{U}) + F_A(\yp{U}) -2  \zeta \yp{U}}{F_A(\yp{U}) - F_{-A}(\yp{U})}.
\end{equation*}
Our nondimensional parameters are always greater than zero, so $0 < 1-e^{\pi_4}$ and \yp{$\zeta<e^{\pi_4}(\pi_6+\zeta)$}. It follows that the derivative of $\hat \phi$ with respect to $\yp{U}$ is nonzero:
\begin{equation*}
\left.\frac{d\hat\phi}{d \yp{U}}\right|_{\yp{U}=0} = \yp{\frac{1}{2}\frac{e^{\pi_4}(\pi_6+\zeta)-\zeta}{e^{\pi_4}-1}\neq 0}.
\end{equation*}
We then obtain a local equation for $\yp{U}$ as a function of $\hat \phi$ by invoking the inverse function theorem:
\begin{equation}
\yp{U}(\hat\phi) = \yp{2\hat\phi \frac{e^{\pi_4}-1}{e^{\pi_4}(\pi_6+\zeta)-\zeta} + O(\hat\phi^2)}.
\end{equation}


We have seen that in some parameter regimes, viscous drag grows to large values during translocation (Figure \ref{fig:c0}C,F), but nonzero velocities persist due to unequal motor ratios (Figure \ref{fig:constriction}C,F). To establish this inverse relationship between velocity and drag, we Taylor expand about infinity with $\hat \zeta = 1/\zeta$ as defined above. Then solving for $\hat \zeta$ in terms of $\yp{U}$ yields
\begin{equation*}
\hat\zeta = \frac{\yp{U}}{\phi F_{-A}(\yp{U}) + (1-\phi)F_A(\yp{U})}.
\end{equation*}
To derive a local equation for $\yp{U}$ as a function of $\hat\zeta$, we examine the derivative of $\hat\zeta$ with respect to $\yp{U}$:
\begin{equation*} 
\left. \frac{d\hat \zeta}{d \yp{U}}\right|_{\yp{U}=0}=\yp{\frac{1}{2\phi-1}}.
\end{equation*}
So long as $\phi \neq 1/2$, this derivative is well-defined, and we can invoke the inverse function theorem to write the local equation for $\yp{U}$ as a function of $\hat\zeta$.
\begin{equation*}
\yp{U}(\hat \zeta) = \yp{\hat \zeta(2\phi-1)} + O(\hat\zeta^2).
\end{equation*}
After a trivial substitution, we arrive at the desired equation,
\begin{equation*}
\yp{U}(\zeta) = \yp{\zeta^{-1}  (2\phi-1)} + O\left(\zeta^{-2}\right) = \yp{2 \zeta^{-1}\hat \phi} + O\left(\zeta^{-2}\right).
\end{equation*}
%
Combining the local velocity estimates, we arrive at the \yp{local} velocity \yp{existence} equation as a function of $\hat\phi$ and $\zeta$:
\begin{equation}\label{eq:V}
\yp{U}(\hat\phi,\zeta) = \yp{2\hat\phi \left[\frac{e^{\pi_4}-1}{e^{\pi_4}(\pi_6+\zeta)-\zeta} + \frac{1}{\zeta}\right] } + O(\hat\phi \zeta^{-1},\hat\phi^2,\zeta^{-2}).
\end{equation}
This existence equation is valid for large $\zeta$ and any $\hat\phi\in[-1/2,1/2]$, or for small $\hat\phi$ and any $\zeta$.

Equation \eqref{eq:V} justifies the effects observed from simulations earlier in the paper. For $\zeta$ large, the velocity must be small and proportional to $\zeta^{-1}$, and the sign of $\hat\phi$ determines the sign of the velocity. It is the ratio of motors that determines the direction of motion for large drag. In the case of small drag, this equation is only valid for small $\hat\phi$, but the equation yields the same intuition that is consistent with our one-parameter bifurcation diagrams: the dominant motor species determine the sign of the velocity.

In the case of large drag, or small drag with near-equal (but non-equal) motor ratios, increasing the velocity scale ratio between translocation and motor adhesion dynamics $\pi_6$ will \yp{decrease} translocation velocity. Finally, velocity depends weakly on the initial motor displacement $\pi_4$ and is entirely independent of the ratio of motor detachment position and the maximum displacement of each motor ($\pi_3$ and $\pi_5$). \yp{Therefore, we expect that even significant changes to the motor displacements or substantial changes to motor attachment and detachment rates will have little effect on translocation dynamics.}

\subsection{Linear Stability} 

For our one-dimensional problem, a linear stability analysis is sufficient to understand the stability of fixed points, which follows from the slope of the total force function:
\begin{equation}\label{eq:stability}
\lambda = F'(\yp{U}) - \zeta.
\end{equation}
Note that if the derivative of $F$ is bounded, it follows that the eigenvalue is negative ($\lambda < 0$) for sufficiently large drag forces. Therefore, all velocities satisfying force-balance are stable. We can rule out multistability for large drag because, if there is more than one fixed point, there must also be more than one unstable fixed point. This is impossible by the negativity of $\lambda$ shown above. Using the continuity of the force-velocity curves (Appendix \ref{a:cont}), the proof follows by contradiction: if there are only two fixed points that are both stable, the slope of the net force function must be negative at each point. By the intermediate value theorem, there must exist a third point between them, which contradicts the original claim of two stable points. So, either one point must be unstable, or there is only one fixed point. This argument can be extended to eliminate any number of stable fixed points in the case of large drag.



In the special case $\yp{U}=0$, the eigenvalue can be computed explicitly to yield
\begin{equation*}
\lambda = \frac{e^{\pi_4}\pi_6}{1-e^{\pi_4}} - \zeta.
\end{equation*}
In the physically relevant parameter regime, the conditions $\pi_4$, $\pi_6>0$, and $\zeta\geq0$ always hold. Therefore $\lambda <0$ for any choice of parameters, implying that the stationary solution is always stable.

For small deviations from $\phi=0.5$, we use Equation \eqref{eq:V} to rewrite the eigenvalue:
\begin{align*}
\lambda &= F'\left(0 + 2\hat\phi \frac{e^{\pi_4}-1}{e^{\pi_4}\pi_6} + O(\hat\phi^2)\right) -  \zeta\\
&=F'(0) + 2\hat\phi \frac{e^{\pi_4}-1}{e^{\pi_4}\pi_6} F''(0) - \zeta + O(\hat\phi^2)\\
&=\frac{e^{\pi_4}\pi_6}{1-e^{\pi_4}} + 2\hat\phi \frac{e^{\pi_4}-1}{e^{\pi_4}\pi_6} \left[ \frac{2 e^{\pi_4}(2\yp{\phi}-1)\pi_6^2}{e^{\pi_4}-1}\right] - \zeta+ O(\hat\phi^2).
\end{align*}
Because we assume that $\yp{\phi} = 0.5 + \hat\phi$, the term $2\yp{\phi}-1$ reduces to $2\hat\phi$, making the second term order $O(\hat\phi^2)$. Therefore, the eigenvalue equation reduces to
\begin{equation*}
\lambda =  \frac{e^{\pi_4}\pi_6}{1-e^{\pi_4}} -\zeta + O(\hat\phi^2).
\end{equation*}
Recalling that $\pi_4,\pi_6>0$, we generally expect constant velocity solutions to be stable for $\hat\phi$ small.

\section{Discussion}

In this paper, we fully characterize the dynamics predicted by a model of vesicles driven into closed constrictions. We cast the system into a fast-slow system, perform a two-parameter bifurcation analysis on the fast subsystem, then determine how the dominant motor species affect the cusps in the resulting bifurcation surface. The model predicts multistability, i.e., bidirectional motion, for smaller values of viscous drag and unidirectional motion for greater values of the viscous drag corresponding to tight constrictions.

We remark that our reduced axisymmetric lubrication model captures the diverse morphology of dendritic spines, from thin spines that are often less than $2$\si{.\um} in length with neck diameters ranging from 0.06--0.2\si{.\um} \cite{arellano2007ultrastructure} to mushroom-shaped spines that are often less than $1$\si{.\um} in width \cite{risher2014rapid}. As long as the diameters of the vesicle and spine wall are similar, our theory applies.

\subsection{Physiological Parameters}
While detailed electron microscopy images exist of dendritic spines \cite{kasthuri2015saturated}, it is a significant challenge to search and classify recycling endosomes efficiently. For now, we rely on published endosome images to approximate the physiologically relevant parameter ranges.

\begin{figure}[ht!]
	\centering
	\includegraphics[width=\textwidth]{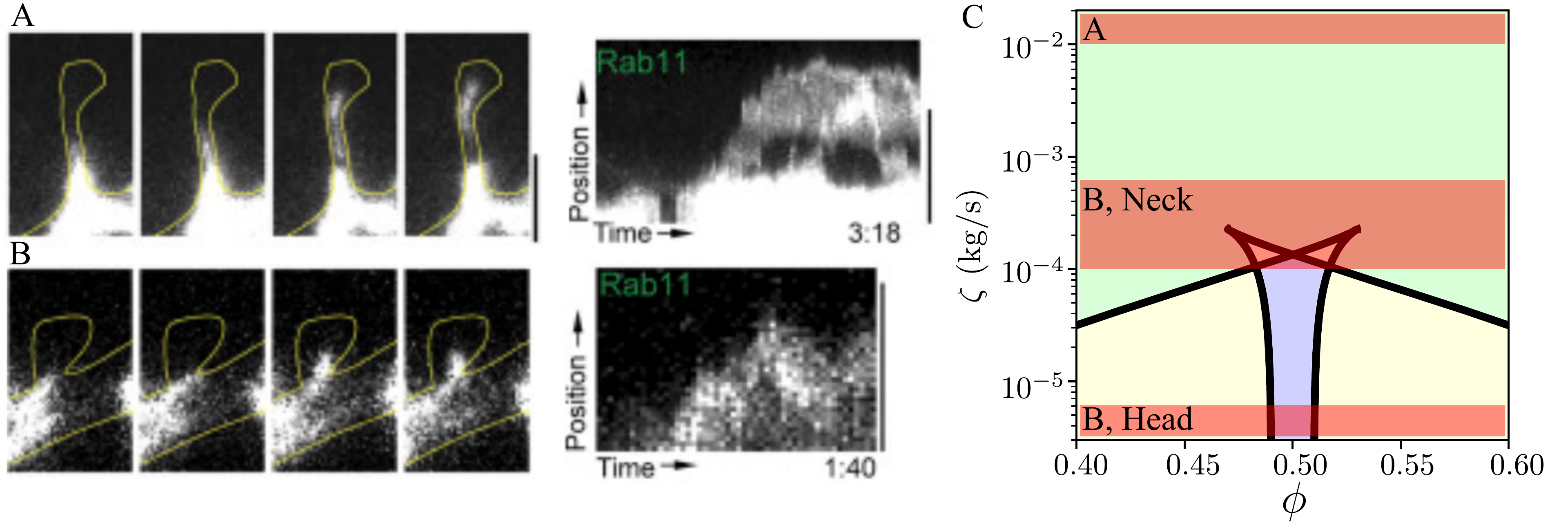}
	\caption{Time-lapse images of recycling endosomes adapted from \cite{da2015positioning} and available under the CC BY NC ND license. A: a recycling endosome translocates through a thin spine in four time-lapse images (left). A kymograph is shown to the right. The vesicle is roughly $1$\si{.\um} in diameter, and the distance between the vesicle and neck wall is at most 0.1\si{.\um}. B: a recycling endosome translocates into a stubby spine in a series of time-lapse images, with the associated kymograph on the right. The vesicle is roughly $0.5$\si{.\um} in diameter, the distance between the vesicle and neck wall is roughly 0.15\si{.\um}, and the distance between the vesicle and spine wall is roughly 0.5\si{.\um}. All scale bars 2\si{.\um}. C: approximate ranges of drag (red transparent) superimposed on the two-parameter diagram from Figure \ref{fig:2par} with the drag $\zeta$ plotted on a log scale. Labels A and B correspond to Panels A and B. }\label{fig:vesicle_example}
\end{figure}

In Figure \ref{fig:vesicle_example}A, B, we show two representative experimental images from \cite{da2015positioning} of spines containing recycling endosomes. A thin yellow line outlines the spine, and the endosome travels through the spine neck in a series of four time-lapse images with an associated kymograph on the right. The scale bar is 2\si{.\um} in all panels. Using these images, we determine approximate regions where physiological parameters may lie in the $\phi$-$\zeta$ parameter space (Figures \ref{fig:2par}D and \ref{fig:vesicle_example}C). Figure \ref{fig:vesicle_example}C is the same as Figure \ref{fig:2par}D but with $\zeta$ plotted on a log scale. The time-lapse images provide virtually no information about the ratio of motors, so we make no restrictions on $\phi$ for now.

We estimate the viscous values from these images by estimating the height between the vesicle and spine wall in Figure \ref{fig:vesicle_example}A and assume a constant constriction for simplicity. Through a crude manual approximation, we estimate the height in panel A to be at most 0.1\si{.\um}. This height is substantially smaller than the heights considered in this paper, and therefore yields a relatively greater viscous drag value of $\zeta\approx$ \si{\num{2e-2} .kg/s} (we assume the same microscopic motor parameters as in Figure \ref{fig:2par}). This value corresponds to a point far above the cusps, where only a single-velocity solution exists (Figure \ref{fig:vesicle_example}C, top red band labeled ``A''). This range lies well within the unidirectional regime. Because the time-lapse images in Figure \ref{fig:vesicle_example}A show the endosome traveling towards the spine head, we infer that the upwards-pushing motors are dominant. The only region where the vesicle could switch direction is at the base of the spine, where there some distance between the diameter of the cell wall and the recycling endosome is possible. Thin spines generally are much smaller in diameter relative to recycling endosomes, so that the distance between the vesicle and spine wall is small. We conclude that once the vesicle has entered a thin spine, we generally expect unidirectional movement.

Figure \ref{fig:vesicle_example}B shows an example of a stubby spine. Here, the endosome is smaller, with a diameter range from $0.5$\si{.\um} to $0.6$\si{.\um}. We estimate that the height between the endosome and neck is roughly 0.05--0.1\si{.\um}, which yields viscous drag values in the range $\zeta\approx$ \si{\num{1e-4}} to \si{\num{6e-4}.kg/s}. Recalling that the cusps in Figure \ref{fig:2par} are on the order of \si{\num{2e-4}.kg/s} (where the critical vesicle-to-spine diameter ratio is roughly 2\si{.\um}/3\si{.\um}), we find that both unidirectional and multidirectional motion are possible (Figure \ref{fig:vesicle_example}C, middle red band labeled ``B, neck''). Unidirectional solutions take a much greater portion of the parameter range on a linear scale. Another property to consider is that stubby spines have shallow constrictions that lead into a large head, where the vesicle may spend a substantial amount of time relative to the neck. Therefore, it is worth considering whether multistable solutions exist past the neck. In the head, the height between the endosome and head ranges from roughly 0.1\si{.\um} to 0.5\si{.\um}, which yields a drag range of \si{\num{3e-6}.kg/s} to \si{\num{1e-4}.kg/s}, which places the vesicle squarely in a multistable regime  (Figure \ref{fig:vesicle_example}C, lower red band labeled ``B, head''). Strikingly, the kymograph shows unidirectional movement through the neck and bidirectional motion in the spine head.

The viscous drag in these representative examples varies over four orders of magnitude, and our model predicts dramatically different qualitative behaviors over this range. According to our model, thin spines with large vesicles will exhibit unidirectional movement due to the large drag experienced by the vesicle. In contrast, stubby spines will exhibit multidirectional motion, especially if the diameter of the recycling endosome is smaller than the spine. Interestingly, published images of vesicle movement in spines appear to confirm these claims.


\subsection{Limitations and Future Directions}\label{sec:limitations}

There are important caveats to the present study. One is the use of mean-field models. In particular, the force-velocity functions used in this paper rely on the limit of large numbers of myosin motors \cite{hoppensteadt2012modeling}, and therefore questions about noise cannot be addressed in this framework. One possible approach to overcome this limitation is to replace the current mean-field model of molecular motors by a discrete model; such a discrete model is used to compute mean passage times in Allard et al. \cite{allard2019bidirectional}. Finally, numerical experiments are another feasible approach to address the question of mean first passage times of a translocating vesicle. Similar approaches have been performed in \cite{dallon2019stochastic} to pursue questions regarding the effects of intermediate filament parameters on intermediate filament transport.

\yp{Another limitation of the present work is the rigidity assumption $\pi_2=0$, which is equivalent to assuming a rigid, undeformable spherical vesicle. In fact, experimental images show that vesicles undergo large deformations as they squeeze through spine necks. One could, in principle, incorporate this into the present model by adding an equation for the vesicle shape based on its elastic properties, surface area to volume ratio, and the surrounding flow. In the present work we have chosen to focus on the simpler case of a spherical vesicle, since in this case the height function may be parameterized in terms of a single geometric parameter, the radius, which simplifies the problem considerably and allows a more straightforward analysis.}

\yp{The insights gained from studying the transport of single vesicles into dendritic spines are expected to apply to the population-level dynamics with some caveats. A potential application is that multistability noted in the context of single vesicles would translate into the relative fractions of vesicles in the population that experience translocation, rejection, and corking. However, one complicating factor is that the dynamics of different vesicles are not independent---they are coupled through the fluid mechanics of the intracellular fluid. This long-ranged fluid-structure interaction is an important subject for future work.}

\section{Acknowledgments}
The authors acknowledge support under the National Institute of Health grant T32 NS007292 (YP) and National Science Foundation grant DMS-1913093 (TGF). \yp{The authors thank Chris H. Rycroft and Jonathan Touboul for reading the early version of the text and providing useful feedback.}

\appendix

\section{Numerics}

In this section, we detail the numerical methods and parameters of the fast-slow lubrication model.

\subsection{Integration}\label{a:integration}

For convenience, we restate the original fast-slow lubrication model here:
\begin{equation}\label{eq:fs1_appendix}
\begin{split}
\frac{dZ}{dt} &= U,\\
\ve\frac{dU}{dt} &= F(U) - \zeta(Z) U.
\end{split}
\end{equation}

We solve Equation \eqref{eq:fs1_appendix} numerically by taking $\ve$ nonzero and integrating forwards in time. \yp{We obtain visually indistinguishable results using either Python's \texttt{odeint} routine or the forward Euler method}. This approach is beneficial because velocities satisfying instantaneous force-balance, i.e., $U$ such that $F(U) - \zeta(Z)U=0$, are fixed points of a one-dimensional ODE (assuming $Z$ is constant). The question of convergence is then straightforward: if an initial condition is within the basin of attraction of a stable velocity, it will converge to this velocity. 

Another benefit of taking this approach is that for $\ve$ sufficiently small, system \eqref{eq:fs1_appendix} is sufficiently similar to the overdamped system when $\ve=0$, as guaranteed by Fenichel theory \cite{fenichel1979geometric,broer2013geometric}). When $\ve\ll 1$, there is a separation of timescales, and very small time steps are required for numerical stability. We find that choosing $\ve=1$ works well in practice, as the dynamics follow the slow manifold but do not require impractically small time steps. This approximation works even at moderate values of $\ve$ (e.g., at $\ve=1$): whereas we have nondimensionalized velocity in terms of the free space prediction of Stokes' law, in confined geometries there is a small parameter which may be obtained by rescaling $\ve$ by the effective drag. For a given choice of $\ve$, we typically take the time step \texttt{dt} to be smaller by one or two orders of magnitude.

\subsection{Continuation}\label{a:continuation}

General continuation strategies in one and two parameters can be found in Chapter 10.3 of Kuznetsov \cite{kuznetsov2013elements}. We use XPPAUTO \cite{xpp} (version 8) to generate our bifurcation diagrams unless stated otherwise. For the generation of the one- and two-parameter diagrams, we use the following numerical values:
\begin{verbatim}
Ntst	= 15 (default)
Nmax	= 200 (default)
Npr		= 50 (default)
Ds		= 1e-10
Dsmin	= 1e-10
Ncol	= 4 (default)
Dsmax	= 0.1
\end{verbatim}
All other numerical values remain at default values. We adjust \texttt{Ntst}, \texttt{Nmax}, \texttt{Npr}, \texttt{Par Min}, and \texttt{Par Max} as needed. The displayed \texttt{Dsmax} value may make AUTO run multiple passes over some branches, in which case we reduce \texttt{Dsmax} to 0.001, preventing this issue while incrementing at a reasonable pace. We include many more details in a mini-tutorial on our \yp{GitHub} repository at \url{https://github.com/youngmp/park_fai_2020}.


\subsection{Cusps}\label{a:cusps}
Saddle-nodes may be computed by finding tangencies in the right-hand side of Eq. \eqref{eq:fast} with the constraint that $U$ satisfies
\begin{equation*}
0 = \phi F_{-A}(U) + (1-\phi) F_A(U) - \zeta U
\end{equation*}
The conditions for a saddle-node bifurcation are \cite{kuznetsov2013elements}:
\begin{equation}\label{eq:sn_conditions}
\begin{split}
f(U) &= F(U,\phi)-\zeta U =0,\\
f'(U) &= F'(U,\phi) -\zeta = 0,\\
f''(U) &= F''(U) \neq 0.
\end{split}
\end{equation}
We denote any $U$ that simultaneously satisfies these equations by $U^*$. We can simplify the search for saddle-nodes by writing the system
\begin{equation}\label{eq:fast_saddles}
\begin{split}
\bar U' &= F(\bar U,\bar \phi) -\zeta \bar U.\\
\bar \phi' &= F'(\bar U,\bar \phi) - \zeta.
\end{split}
\end{equation}
We can just as easily use $\bar \zeta$ and look for fixed points in $(\bar U, \bar\zeta)$ as a function of $\phi$. It is merely a matter of preference. We use $\bar U$ and $\bar \phi$ to emphasize the difference between \eqref{eq:fast_saddles} and the original fast subsystem \eqref{eq:fast}. This new system eliminates one parameter when it comes to computing bifurcation diagrams. Fixed points of \eqref{eq:fast_saddles} correspond to saddle-node bifurcations in the fast subsystem \eqref{eq:fast}. The $\bar U$ nullcline of \eqref{eq:fast_saddles} shows the stable fixed points in \eqref{eq:fast}, and the $\bar U$ nullcline intersections with the $\bar \phi$ nullcline correspond to saddle-nodes. Thus, the phase space of \eqref{eq:fast_saddles} corresponds to one-parameter bifurcations in \eqref{eq:fast}, and the one-parameter bifurcations in \eqref{eq:fast_saddles} correspond to two-parameter bifurcations in \eqref{eq:fast}. In particular, saddle-node bifurcations of \eqref{eq:fast_saddles} correspond to cusp bifurcations in \eqref{eq:fast}. We exploit the latter fact to generate the diagram of cusp bifurcations (Figure \ref{fig:cusps}).

In practice, we compute intersections in the nullclines of Equation \eqref{eq:fast_saddles} and track where the intersections change in number. Suppose that we are interested in finding cusp bifurcations as a function of $\pi_4$ and $\pi_5$ for a given $\zeta$. On one side of the cusp bifurcation, Equation \eqref{eq:fast_saddles} shows two fixed points, and on the other side shows zero. By defining a function that returns $+1$ on one side of the bifurcation and $-1$ on the other, we can use a root-finding method such as Brentq to determine $\pi_5$ where there is a transition in the fixed point number to high numerical precision.

\section{Properties of the force-velocity curve}
The force-velocity curves are very well-behaved, but it is not always obvious how. In this section, we briefly show some properties used in the text.

\subsection{Continuity}\label{a:cont}
The continuity of the force-velocity curve (Equation \eqref{eq:fv}) is not immediately apparent. We rewrite the force-velocity curve verbatim for convenience:
\begin{equation}\label{eq:fva} 
\widetilde F_A = \left\{ \begin{matrix}
-\frac{1+\pi_6\widetilde U(e^{\pi_4}-1)^{-1}}{1-\pi_6\widetilde U}, & \text{if}\quad \widetilde U < 0\\
\frac{-(\pi_3+1)}{\pi_3(1-e^{-\pi_5/\pi_6 \widetilde U}) + 1} \frac{[e^{\pi_4}(1-e^{\pi_5}e^{-\pi_5/\pi_6\widetilde U})]-(1-\pi_6\widetilde U)(1- e^{-\pi_5/\pi_6 \widetilde U})}{(e^{\pi_4}-1)(1-\pi_6\widetilde U)}, & \text{if} \quad \widetilde U \geq 0.
\end{matrix}
\right.
\end{equation}
In particular, there are two possible problem areas: when $\widetilde U=0$ and when $\widetilde U=1/\pi_6$. The latter case is especially important because numerical problems occasionally arise when evaluating $\wu=1/\pi_6$ directly. In the first case, the left limit is straightforward:
\begin{equation}
\lim_{\widetilde U\rightarrow 0^-} \wf_{A} =-1.
\end{equation}
The right limit depends on the behavior of the exponentials. Noting that $\exp(-\pi_5/\pi_6\wu) \rightarrow 0$ as $\wu \rightarrow 0^+$, it is straightforward to check that the left and right limits agree independent of the parameters $\pi_i$, and therefore $\wf_A$ is continuous at $\wu=0$. Continuity of $\wf_{-A}$ follows by definition.

In the second case, when $\wu =1/\pi_6\geq0$, we take a close look at the second line of Equation \eqref{eq:fva}. Potential problems arise in the term $(1-e^{\pi_5}e^{-\pi_5/\pi_6\widetilde U})/(1-\pi_6\widetilde U)$. For convenience, let $v:=1-1/(\pi_6 \wu)$ so that $v\rightarrow0$ as $\wu\rightarrow1/\pi_6$. Then the term becomes
\begin{align*}
\frac{1-e^{\pi_5}e^{-\pi_5/\pi_6\widetilde U})}{1-\pi_6\widetilde U} &=-\frac{(1-e^{\pi_5v})(1-v)}{v}\\
&=-\frac{[1-(1+\pi_5v + O(v^2))](1-v)}{v}\\
&=-\frac{[1-(1+v(\pi_5-1) + O(v^2))]}{v}\\
&=\frac{v(\pi_5-1) + O(v^2))}{v}\\
&=\pi_5-1 + O(v).
\end{align*}
We have used a Taylor expansion of the exponential about zero. So the term converges to $\pi_5-1$ as $v\rightarrow0$ on either side of the limit. It follows that the function $\wf_A$ is continuous at $\wu=1/\pi_6$, as desired.



\subsection{Limits}

Consider $\pi_4\rightarrow 0$. This limit occurs when attachments occur at zero position ($A=0$), or when the force-scaling parameter $\gamma=0$. Both cases appear to be somewhat unrealistic: in the former case, newly attached crossbridges will apply no force (Equation \eqref{eq:motor_force}), and in the latter case, the force exerted by a single motor is always zero. We find that the term $e^{\pi_4}-1$ appears in the denominator of several terms and competes with $U$ when $U$ is small. For $|U|\gg0$, the term $(e^{\pi_4}-1)^{-1}$ dominates, and we expect $F_A(U)\rightarrow \infty$ as $\pi_4 \rightarrow 0$. When $|U|\ll1$, we expect $F_A(U)\rightarrow -1$ as $U \rightarrow 0$ for each $\pi_4$ small.
\begin{equation*}
F_A( U,\pi_4\rightarrow 0) = \left\{ \begin{matrix}
\infty &  U < 0,\\
-1 &  U = 0,\\
-\infty &  U \geq 0.
\end{matrix} \right.
\end{equation*}
In contrast, the dimensional force-velocity curve converges for $\pi_4$ small and diverges for $\pi_4$. A nondimensionalization using $F_1 = F_0/(e^{\pi_4}-1)$ would remove this problem of divergence, but this rescaling is not necessary when considering bifurcations: for any fixed point $F(U,\pi_4)=0$, where $F$ is the nondimensional net force, multiplying both sides by the scaling factor $(e^{\pi_4}-1)$ yields $(e^{\pi_4}-1)F(U,\pi_4)=0$, so scaling preserves fixed points as a function of $\pi_4$.
\yp{
	\section{Comparisons to the Allard et al. (2019) Symmetric Kinesin Motor Model}
	
	The choice of motor model does not qualitatively change multistability as a function of drag. Using a similar symmetric motor model of kinesin motors from \cite{allard2019bidirectional}, we add an analogous constriction term in their force-balance equation and rederive the master equation. To make direct comparisons to our results, we go a step further and derive the Fokker-Planck equation, from which we produce bifurcation diagrams. To minimize confusion, we rephrase their problem in terms of molecular motors in spines. In particular, we consider two identical species of kinesin motors, where one species prefers to push up and the other prefers to push down. 
	
	We note several important differences that allows a very straightforward derivation of the mean-field equations in the symmetric kinesin model. first, in contrast to the myosin motors considered in the present study, the kinesin motor model does not have spatial dependence. Second, in contrast the the attachment and detachment rate of myosin motors that depend on position-dependent rates, the attachment and detachment rates depend continuously on the number of attached motors. Third, the kinesin motor model requires that a total of $K$ motors are always attached. For example, if there are $N$ up motors attached to the vesicle, then there must be $M=K-N$ down motors attached. Importantly, we derive the motor detachment rates and assume instantaneous reattachment of either species. We now proceed with the derivation.
	
	In the symmetric kinesin model, the forces exerted in the up and down directions are given by
	\begin{equation*}
	f^+ = F_s\left( 1 - \frac{U}{V_m}\right), \quad \quad f^- = F_s\left(-1 - \frac{U}{V_m} \right),
	\end{equation*}
	where $F_s$ is the motor stall force, $U$ is the cargo velocity, and $V_m$ is the free-moving velocity of a single motor. Instantaneous force-balance requires that 
	\begin{equation}\label{eq:fb_allard}
	F_sN\left( 1 - \frac{U}{V_m}\right) + F_sM\left(-1 - \frac{U}{V_m} \right) - \zeta U = 0,
	\end{equation}
	where $N$ and $M$ are the number of attached motors pushing in the up and down directions, respectively. Note that this equation is the same starting point considered by \cite{allard2019bidirectional}, but we have included a drag force $\zeta U$. Moreover, this equation is the discrete analogue of our force-balance equation, $\phi F_{-A}(U) + (1-\phi) F_A(U) -\zeta U = 0$. In contrast to our formulation, it is possible to solve explicitly for the velocity. Solving for $U$ yields,
	\begin{equation}\label{eq:u_allard}
	U = \frac{N-M}{N/V_m + M/V_m + \zeta/F_s}.
	\end{equation}
	Plugging into Equation \eqref{eq:fb_allard} allows us to write the up and down forces in terms of the total motors attached:
	\begin{equation}\label{eq:forces}
	f^+(M,N) = F_s \left(1-\frac{N-M}{N+M+ \zeta V_m/F_s} \right), \quad f^-(M,N) = F_s \left( 1 + \frac{N-M}{N+M+\zeta V_m/F_s}\right).
	\end{equation}
	Attachment and detachment rates follow from Bell's law \cite{bell1978models}:
	\begin{align}\label{eq:xi}
	\xi^+(M,N) = N\bar \kappa_0 \exp\left(\frac{f^+(M,N)}{f_0}\right), \quad\quad \xi^-(M,N) = M\bar \kappa_0 \exp\left(\frac{f^-(M,N)}{f_0}\right).
	\end{align}
	Plugging Equation \eqref{eq:forces} into Equation \eqref{eq:xi}, the attachment and detachment rates become
	\begin{align*}
	\kappa_M^+ &= \frac{\bar \kappa_0}{2}(K-M)  \exp\left(\frac{F_s}{f_0} \left(1 + \frac{2M - K}{K + \zeta V_m/F_s}\right)\right),\\
	\kappa_M^- &= \frac{\bar \kappa_0}{2}M \exp\left( \frac{F_s}{f_0} \left(1 - \frac{2M-K}{K+ \zeta V_m/F_s}\right)\right),
	\end{align*}
	where we take $M=K-N$. To nondimensionalize, we take $\gamma = 2F_s/f_0$, $\tilde\zeta = \zeta V_m/(2 F_s)$, and $\kappa_0 = \bar\kappa_0/2$. The nondimensional parameters yield,
	\begin{align*}
	\kappa_M^+ &= \kappa_0(K-M)  \exp \left( \gamma \left(\frac{M + \tilde \zeta}{K + 2\tilde\zeta}\right)\right),\\
	\kappa_M^- &= \kappa_0 M \exp\left(\gamma\left(\frac{ K-M + \tilde \zeta }{K+ 2\tilde \zeta}\right)\right).
	\end{align*}
	where $\gamma = 2F_s/f_0$, $\tilde\zeta = \zeta V_m/(2 F_s)$, and $\kappa_0 = \bar\kappa_0/2$. The $\gamma$ and $\kappa_0$ parameters are identical to the parameters used in Allard. The $\tilde\zeta$ term is the nondimensionalized drag.
	\begin{figure}[ht!]
		\makebox[\textwidth][c]{
			\centering
			\includegraphics[width=.75\textwidth]{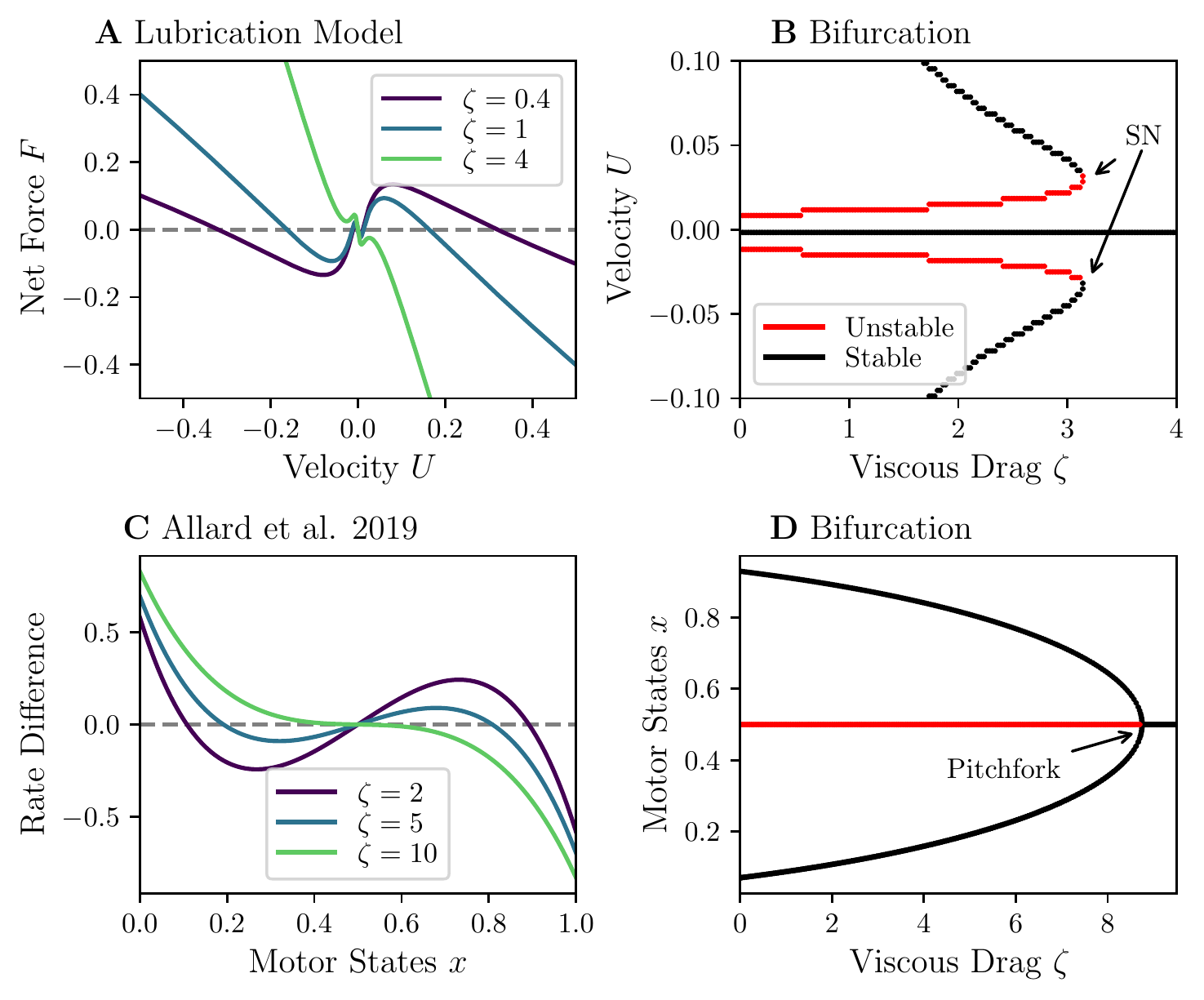}}
		\caption{\yp{Figure C.1: A comparison of bifurcations in the lubrication and symmetric kinesin models. A: Example force-velocity curves from the lubrication model with different values of viscous drag. Here we use the parameters $\pi_1=1$, $\pi_3=1$, $\pi_4=4.7$, $\pi_5=0.1$, $\phi=0.5$. B: Corresponding one-parameter bifurcation. Multistability terminates through saddle-node (SN) bifurcations C: Phase line curves in the symmetric kinesin model. D: One-parameter bifurcation diagram of the symmetric kinesin model in $\zeta$. For moderate values of viscous drag $\zeta$, there are two stable motor states. As the drag increases, the system undergoes a pitchfork bifurcation. Beyond the pitchfork, only equal numbers of motors from both species are attached, and therefore only the zero velocity is stable. The parameters used in the kinesin motor model are $\kappa_0=0.5$, $\gamma=3$, $K=35$.}}\label{fig:allard}
	\end{figure}
	We abuse notation and define $\zeta \equiv \tilde\zeta$. The continuum limit version of these equations is taken by noting that
	\begin{equation*}
	\frac{M+\zeta}{K + 2\zeta} = \frac{1}{1+2\zeta/K} \frac{M}{K} + \frac{\zeta}{K + 2\zeta},
	\end{equation*}
	i.e., viscous drag modifies the continuum variable $x:=M/K$ by changing the slope and y-intercept of the input to the exponential functions. Thus, the continuum approximation is given by
	\begin{equation*}
	\kappa_M^+(x) = \kappa_0 (1-x)\exp\left(\gamma(a x + b)\right), \quad\quad \kappa_i^-(x) = \kappa_0 x \exp\left(\gamma(a(1-x)+b)\right),
	\end{equation*}
	where $a=1/(1+2\zeta/K)$, $b=\zeta/(K+2\zeta)$. Using standard methods \cite{gardiner1985handbook}, the deterministic equation is
	\begin{equation*}
	\frac{dx}{dt} = \kappa_M^+(x)-\kappa_M^-(x).
	\end{equation*}
	Figure \ref{fig:allard}B shows the phase line of the deterministic dynamics for different values of $\zeta$. Note that stable motor states inform us directly of stable velocities using Equation \eqref{eq:u_allard}. The magnitude and sign of the velocity is determined by the difference in the number of attached and unattached motors. For example, if the motor states $x=0.1$ and $x=0.9$ are stable, these points correspond to distinct negative and positive velocities. Therefore, we use Figure \ref{fig:allard}B as a proxy for determining the existence of multistable velocities in the symmetric kinesin model.
	
	We include the lubrication force-velocity curves in Figure \ref{fig:allard}A and its corresponding bifurcation diagram in panel B as a reminder of how multistability changes as a function of drag. For this comparison, we can only take $\phi=0.5$. Recall that in the lubrication model, nontrivial velocities exist for a range of $\zeta$ up to the saddle-node (SN) bifurcation. The stable non-zero velocities terminate through a pair of saddle-node bifurcations and leave one stable zero velocity.
	
	Figure \ref{fig:allard}C shows a similar a one-parameter diagram of the deterministic symmetric kinesin model. For a range of $\zeta$ up to the pitchfork bifurcation, two different motor states, and therefore two different velocities are stable. Beyond the pitchfork, only one stable zero velocity solution persists.
	
	In summary, the specifics of the lubrication and symmetric kinesin models differ substantially. The symmetric kinesin models do not take into account motor head positions, and have attachment and detachment rates dependent on the number of different species attached. In contrast, the myosin model includes position-dependent forces, and it is this position-dependence that gives rise to local extrema near zero velocity in the myosin force-velocity curves. Thus, the two models lose multistability through different bifurcations. However, the broad qualitative behaviors are similar: multistable cargo velocities exist for moderate values of $\zeta$, and sufficiently large drag values cause the cargo to stop moving.

}
\bibliographystyle{plain}

\begin{thebibliography}{46}
	\providecommand{\natexlab}[1]{#1}
	\providecommand{\url}[1]{\texttt{#1}}
	\expandafter\ifx\csname urlstyle\endcsname\relax
	\providecommand{\doi}[1]{doi: #1}\else
	\providecommand{\doi}{doi: \begingroup \urlstyle{rm}\Url}\fi
	
	\bibitem[Acheson(1991)]{acheson1991elementary}
	David~J Acheson.
	\newblock Elementary fluid dynamics, 1991.
	
	\bibitem[Adrian et~al.(2014)Adrian, Kusters, Wierenga, Storm, Hoogenraad, and
	Kapitein]{adrian2014barriers}
	Max Adrian, Remy Kusters, Corette~J Wierenga, Cornelis Storm, Casper~C
	Hoogenraad, and Lukas~C Kapitein.
	\newblock Barriers in the brain: resolving dendritic spine morphology and
	compartmentalization.
	\newblock \emph{Frontiers in Neuroanatomy}, 8:\penalty0 142, 2014.
	
	\bibitem[Allard et~al.(2019)Allard, Doumic, Mogilner, and
	Oelz]{allard2019bidirectional}
	Jun Allard, Marie Doumic, Alex Mogilner, and Dietmar Oelz.
	\newblock Bidirectional sliding of two parallel microtubules generated by
	multiple identical motors.
	\newblock \emph{Journal of Mathematical Biology}, pages 1--24, 2019.
	
	\bibitem[Arellano et~al.(2007)Arellano, Benavides-Piccione, DeFelipe, and
	Yuste]{arellano2007ultrastructure}
	Jon~I Arellano, Ruth Benavides-Piccione, Javier DeFelipe, and Rafael Yuste.
	\newblock Ultrastructure of dendritic spines: correlation between synaptic and
	spine morphologies.
	\newblock \emph{Frontiers in Neuroscience}, 1:\penalty0 10, 2007.
	
	\bibitem[Bagnall et~al.(2015)Bagnall, Byun, Begum, Miyamoto, Hecht, Maheswaran,
	Stott, Toner, Hynes, and Manalis]{bagnall2015deformability}
	Josephine~Shaw Bagnall, Sangwon Byun, Shahinoor Begum, David~T Miyamoto,
	Vivian~C Hecht, Shyamala Maheswaran, Shannon~L Stott, Mehmet Toner, Richard~O
	Hynes, and Scott~R Manalis.
	\newblock Deformability of tumor cells versus blood cells.
	\newblock \emph{Scientific Reports}, 5:\penalty0 18542, 2015.
	
	\bibitem[Bell(1978)]{bell1978models}
	George~I Bell.
	\newblock Models for the specific adhesion of cells to cells.
	\newblock \emph{Science}, 200\penalty0 (4342):\penalty0 618--627, 1978.
	
	\bibitem[Berger et~al.(2018)Berger, Seung, and Lichtman]{berger2018vast}
	Daniel~R Berger, H~Sebastian Seung, and Jeff~W Lichtman.
	\newblock Vast (volume annotation and segmentation tool): efficient manual and
	semi-automatic labeling of large 3d image stacks.
	\newblock \emph{Frontiers in Neural Circuits}, 12:\penalty0 88, 2018.
	
	\bibitem[Bressloff and Newby(2009)]{bressloff2009directed}
	Paul Bressloff and Jay Newby.
	\newblock Directed intermittent search for hidden targets.
	\newblock \emph{New Journal of Physics}, 11\penalty0 (2):\penalty0 023033,
	2009.
	
	\bibitem[Bressloff and Newby(2013)]{bressloff2013metastability}
	Paul~C Bressloff and Jay~M Newby.
	\newblock Metastability in a stochastic neural network modeled as a velocity
	jump markov process.
	\newblock \emph{SIAM Journal on Applied Dynamical Systems}, 12\penalty0
	(3):\penalty0 1394--1435, 2013.
	
	\bibitem[Broer et~al.(2013)Broer, Kaper, and Krupa]{broer2013geometric}
	Henk~W Broer, Tasso~J Kaper, and Martin Krupa.
	\newblock Geometric desingularization of a cusp singularity in slow--fast
	systems with applications to zeeman’s examples.
	\newblock \emph{Journal of Dynamics and Differential Equations}, 25\penalty0
	(4):\penalty0 925--958, 2013.
	
	\bibitem[Byun et~al.(2013)Byun, Son, Amodei, Cermak, Shaw, Kang, Hecht,
	Winslow, Jacks, Mallick, and Manalis]{byun2013characterizing}
	Sangwon Byun, Sungmin Son, Dario Amodei, Nathan Cermak, Josephine Shaw, Joon~Ho
	Kang, Vivian~C. Hecht, Monte~M. Winslow, Tyler Jacks, Parag Mallick, and
	Scott~R. Manalis.
	\newblock Characterizing deformability and surface friction of cancer cells.
	\newblock \emph{Proceedings of the National Academy of Sciences}, 110\penalty0
	(19):\penalty0 7580--7585, 2013.
	\newblock ISSN 0027-8424.
	\newblock \doi{10.1073/pnas.1218806110}.
	\newblock URL \url{https://www.pnas.org/content/110/19/7580}.
	
	\bibitem[da~Silva et~al.(2015)da~Silva, Adrian, Sch{\"a}tzle, Lipka, Watanabe,
	Cho, Futai, Wierenga, Kapitein, and Hoogenraad]{da2015positioning}
	Marta~Esteves da~Silva, Max Adrian, Philipp Sch{\"a}tzle, Joanna Lipka, Takuya
	Watanabe, Sukhee Cho, Kensuke Futai, Corette~J Wierenga, Lukas~C Kapitein,
	and Casper~C Hoogenraad.
	\newblock Positioning of ampa receptor-containing endosomes regulates synapse
	architecture.
	\newblock \emph{Cell Reports}, 13\penalty0 (5):\penalty0 933--943, 2015.
	
	\bibitem[Dallon et~al.(2019)Dallon, Leduc, Etienne-Manneville, and
	Portet]{dallon2019stochastic}
	JC~Dallon, C{\'e}cile Leduc, Sandrine Etienne-Manneville, and St{\'e}phanie
	Portet.
	\newblock Stochastic modeling reveals how motor protein and filament properties
	affect intermediate filament transport.
	\newblock \emph{Journal of Theoretical Biology}, 464:\penalty0 132--148, 2019.
	
	\bibitem[Dawson et~al.(2015)Dawson, H{\"a}ner, and Juel]{dawson2015extreme}
	Geoffrey Dawson, Edgar H{\"a}ner, and Anne Juel.
	\newblock Extreme deformation of capsules and bubbles flowing through a
	localised constriction.
	\newblock \emph{Procedia IUTAM}, 16:\penalty0 22--32, 2015.
	
	\bibitem[Duncanson et~al.(2015)Duncanson, Kodger, Babaee, Gonzalez, Weitz, and
	Bertoldi]{duncanson2015microfluidic}
	Wynter~J Duncanson, Thomas~E Kodger, Sahab Babaee, Grant Gonzalez, David~A
	Weitz, and Katia Bertoldi.
	\newblock Microfluidic fabrication and micromechanics of permeable and
	impermeable elastomeric microbubbles.
	\newblock \emph{Langmuir}, 31\penalty0 (11):\penalty0 3489--3493, 2015.
	
	\bibitem[Ermentrout(2002)]{xpp}
	G.~Bard Ermentrout.
	\newblock \emph{{Simulating, analyzing, and animating dynamical systems: a
			guide to {XPPAUT} for researchers and students}}, volume~14.
	\newblock SIAM, 2002.
	
	\bibitem[Fai et~al.(2017)Fai, Kusters, Harting, Rycroft, and
	Mahadevan]{fai2017active}
	Thomas~G Fai, Remy Kusters, Jens Harting, Chris~H Rycroft, and L~Mahadevan.
	\newblock Active elastohydrodynamics of vesicles in narrow blind constrictions.
	\newblock \emph{Physical Review Fluids}, 2\penalty0 (11):\penalty0 113601,
	2017.
	
	\bibitem[Fenichel(1979)]{fenichel1979geometric}
	Neil Fenichel.
	\newblock Geometric singular perturbation theory for ordinary differential
	equations.
	\newblock \emph{Journal of Differential Equations}, 31\penalty0 (1):\penalty0
	53--98, 1979.
	
	\bibitem[Gabriele et~al.(2010)Gabriele, Versaevel, Preira, and
	Th{\'e}odoly]{gabriele2010simple}
	Sylvain Gabriele, Marie Versaevel, Pascal Preira, and Olivier Th{\'e}odoly.
	\newblock A simple microfluidic method to select, isolate, and manipulate
	single-cells in mechanical and biochemical assays.
	\newblock \emph{Lab on a Chip}, 10\penalty0 (11):\penalty0 1459--1467, 2010.
	
	\bibitem[Gardiner(2009)]{gardiner1985handbook}
	C.~W. Gardiner.
	\newblock \emph{Stochastic methods : a handbook for the natural and social
		sciences}.
	\newblock Springer, Berlin, 2009.
	\newblock ISBN 978-3-540-70712-7.
	
	\bibitem[Gray(1959)]{gray1959axo}
	Edward~G Gray.
	\newblock Axo-somatic and axo-dendritic synapses of the cerebral cortex: an
	electron microscope study.
	\newblock \emph{Journal of Anatomy}, 93\penalty0 (Pt 4):\penalty0 420, 1959.
	
	\bibitem[Gu{\'e}rin et~al.(2011)Gu{\'e}rin, Prost, and
	Joanny]{guerin2011motion}
	Thomas Gu{\'e}rin, J~Prost, and J-F Joanny.
	\newblock Motion reversal of molecular motor assemblies due to weak noise.
	\newblock \emph{Physical Review Letters}, 106\penalty0 (6):\penalty0 068101,
	2011.
	
	\bibitem[Holtmaat and Svoboda(2009)]{holtmaat2009experience}
	Anthony Holtmaat and Karel Svoboda.
	\newblock Experience-dependent structural synaptic plasticity in the mammalian
	brain.
	\newblock \emph{Nature Reviews Neuroscience}, 10\penalty0 (9):\penalty0
	647--658, 2009.
	
	\bibitem[Hoppensteadt and Peskin(2012)]{hoppensteadt2012modeling}
	Frank~C Hoppensteadt and Charles~S Peskin.
	\newblock \emph{Modeling and simulation in medicine and the life sciences},
	volume~10.
	\newblock Springer Science \& Business Media, 2012.
	
	\bibitem[J{\"u}licher and Prost(1995)]{julicher1995cooperative}
	Frank J{\"u}licher and Jacques Prost.
	\newblock Cooperative molecular motors.
	\newblock \emph{Physical Review Letters}, 75\penalty0 (13):\penalty0 2618,
	1995.
	
	\bibitem[Kasai et~al.(2010)Kasai, Fukuda, Watanabe, Hayashi-Takagi, and
	Noguchi]{kasai2010structural}
	Haruo Kasai, Masahiro Fukuda, Satoshi Watanabe, Akiko Hayashi-Takagi, and Jun
	Noguchi.
	\newblock Structural dynamics of dendritic spines in memory and cognition.
	\newblock \emph{Trends in Neurosciences}, 33\penalty0 (3):\penalty0 121--129,
	2010.
	
	\bibitem[Kasthuri et~al.(2015)Kasthuri, Hayworth, Berger, Schalek, Conchello,
	Knowles-Barley, Lee, V{\'a}zquez-Reina, Kaynig, Jones,
	et~al.]{kasthuri2015saturated}
	Narayanan Kasthuri, Kenneth~Jeffrey Hayworth, Daniel~Raimund Berger,
	Richard~Lee Schalek, Jos{\'e}~Angel Conchello, Seymour Knowles-Barley, Dongil
	Lee, Amelio V{\'a}zquez-Reina, Verena Kaynig, Thouis~Raymond Jones, et~al.
	\newblock Saturated reconstruction of a volume of neocortex.
	\newblock \emph{Cell}, 162\penalty0 (3):\penalty0 648--661, 2015.
	
	\bibitem[Koehnle and Brown(1999)]{koehnle1999slow}
	Thomas~J Koehnle and Anthony Brown.
	\newblock Slow axonal transport of neurofilament protein in cultured neurons.
	\newblock \emph{The Journal of Cell Biology}, 144\penalty0 (3):\penalty0
	447--458, 1999.
	
	\bibitem[Kunwar et~al.(2011)Kunwar, Tripathy, Xu, Mattson, Anand, Sigua,
	Vershinin, McKenney, Clare, Mogilner, et~al.]{kunwar2011mechanical}
	Ambarish Kunwar, Suvranta~K Tripathy, Jing Xu, Michelle~K Mattson, Preetha
	Anand, Roby Sigua, Michael Vershinin, Richard~J McKenney, C~Yu Clare,
	Alexander Mogilner, et~al.
	\newblock Mechanical stochastic tug-of-war models cannot explain bidirectional
	lipid-droplet transport.
	\newblock \emph{Proceedings of the National Academy of Sciences}, 108\penalty0
	(47):\penalty0 18960--18965, 2011.
	
	\bibitem[Kusters et~al.(2014)Kusters, van~der Heijden, Kaoui, Harting, and
	Storm]{kusters2014forced}
	Remy Kusters, Thijs van~der Heijden, Badr Kaoui, Jens Harting, and Cornelis
	Storm.
	\newblock Forced transport of deformable containers through narrow
	constrictions.
	\newblock \emph{Physical Review E}, 90\penalty0 (3):\penalty0 033006, 2014.
	
	\bibitem[Kuznetsov(2013)]{kuznetsov2013elements}
	Yuri~A Kuznetsov.
	\newblock \emph{Elements of applied bifurcation theory}, volume 112.
	\newblock Springer Science \& Business Media, 2013.
	
	\bibitem[Li et~al.(2015)Li, Sar{\i}yer, Ramachandran, Panyukov, Rubinstein, and
	Kumacheva]{li2015universal}
	Yang Li, Ozan~S Sar{\i}yer, Arun Ramachandran, Sergey Panyukov, Michael
	Rubinstein, and Eugenia Kumacheva.
	\newblock Universal behavior of hydrogels confined to narrow capillaries.
	\newblock \emph{Scientific Reports}, 5:\penalty0 17017, 2015.
	
	\bibitem[M{\"u}ller et~al.(2008)M{\"u}ller, Klumpp, and
	Lipowsky]{muller2008tug}
	Melanie~JI M{\"u}ller, Stefan Klumpp, and Reinhard Lipowsky.
	\newblock Tug-of-war as a cooperative mechanism for bidirectional cargo
	transport by molecular motors.
	\newblock \emph{Proceedings of the National Academy of Sciences}, 105\penalty0
	(12):\penalty0 4609--4614, 2008.
	
	\bibitem[Newby and Bressloff(2010)]{newby2010random}
	Jay Newby and Paul~C Bressloff.
	\newblock Random intermittent search and the tug-of-war model of motor-driven
	transport.
	\newblock \emph{Journal of Statistical Mechanics: Theory and Experiment},
	2010\penalty0 (04):\penalty0 P04014, 2010.
	
	\bibitem[Newby and Bressloff(2009)]{newby2009directed}
	Jay~M Newby and Paul~C Bressloff.
	\newblock Directed intermittent search for a hidden target on a dendritic tree.
	\newblock \emph{Physical Review E}, 80\penalty0 (2):\penalty0 021913, 2009.
	
	\bibitem[Newby and Keener(2011)]{newby2011asymptotic}
	Jay~M Newby and James~P Keener.
	\newblock An asymptotic analysis of the spatially inhomogeneous velocity-jump
	process.
	\newblock \emph{Multiscale Modeling \& Simulation}, 9\penalty0 (2):\penalty0
	735--765, 2011.
	
	\bibitem[Nimchinsky et~al.(2002)Nimchinsky, Sabatini, and
	Svoboda]{nimchinsky2002structure}
	Esther~A Nimchinsky, Bernardo~L Sabatini, and Karel Svoboda.
	\newblock Structure and function of dendritic spines.
	\newblock \emph{Annual Review of Physiology}, 64\penalty0 (1):\penalty0
	313--353, 2002.
	
	\bibitem[Park et~al.(2006)Park, Salgado, Ostroff, Helton, Robinson, Harris, and
	Ehlers]{park2006plasticity}
	Mikyoung Park, Jennifer~M Salgado, Linnaea Ostroff, Thomas~D Helton,
	Camenzind~G Robinson, Kristen~M Harris, and Michael~D Ehlers.
	\newblock Plasticity-induced growth of dendritic spines by exocytic trafficking
	from recycling endosomes.
	\newblock \emph{Neuron}, 52\penalty0 (5):\penalty0 817--830, 2006.
	
	\bibitem[Penzes et~al.(2011)Penzes, Cahill, Jones, VanLeeuwen, and
	Woolfrey]{penzes2011dendritic}
	Peter Penzes, Michael~E Cahill, Kelly~A Jones, Jon-Eric VanLeeuwen, and Kevin~M
	Woolfrey.
	\newblock Dendritic spine pathology in neuropsychiatric disorders.
	\newblock \emph{Nature Neuroscience}, 14\penalty0 (3):\penalty0 285, 2011.
	
	\bibitem[Portet et~al.(2019)Portet, Leduc, Etienne-Manneville, and
	Dallon]{portet2019deciphering}
	St{\'e}phanie Portet, C{\'e}cile Leduc, Sandrine Etienne-Manneville, and John
	Dallon.
	\newblock Deciphering the transport of elastic filaments by antagonistic motor
	proteins.
	\newblock \emph{Physical Review E}, 99\penalty0 (4):\penalty0 042414, 2019.
	
	\bibitem[Risher et~al.(2014)Risher, Ustunkaya, Alvarado, and
	Eroglu]{risher2014rapid}
	W~Christopher Risher, Tuna Ustunkaya, Jonnathan~Singh Alvarado, and Cagla
	Eroglu.
	\newblock Rapid {G}olgi analysis method for efficient and unbiased
	classification of dendritic spines.
	\newblock \emph{PloS One}, 9\penalty0 (9), 2014.
	
	\bibitem[Walker et~al.(2019)Walker, Uchida, Li, Trivedi, Fenn, Monsma,
	Larivi{\'e}re, Julien, Jung, and Brown]{walker2019local}
	Cynthia~L Walker, Atsuko Uchida, Yinyun Li, Niraj Trivedi, J~Daniel Fenn,
	Paula~C Monsma, Roxanne~C Larivi{\'e}re, Jean-Pierre Julien, Peter Jung, and
	Anthony Brown.
	\newblock Local acceleration of neurofilament transport at nodes of {R}anvier.
	\newblock \emph{Journal of Neuroscience}, 39\penalty0 (4):\penalty0 663--677,
	2019.
	
	\bibitem[Wang et~al.(2008)Wang, Edwards, Riley, Provance, Karcher, Li, Davison,
	Ikebe, Mercer, Kauer, and Ehlers]{wang2008myosin}
	Zhiping Wang, Jeffrey~G. Edwards, Nathan Riley, D.~William Provance, Ryan
	Karcher, Xiang{-}dong Li, Ian~G. Davison, Mitsuo Ikebe, John~A. Mercer,
	Julie~A. Kauer, and Michael~D. Ehlers.
	\newblock Myosin {V}b mobilizes recycling endosomes and ampa receptors for
	postsynaptic plasticity.
	\newblock \emph{Cell}, 135\penalty0 (3):\penalty0 535 -- 548, 2008.
	\newblock ISSN 0092-8674.
	\newblock \doi{https://doi.org/10.1016/j.cell.2008.09.057}.
	\newblock URL
	\url{http://www.sciencedirect.com/science/article/pii/S0092867408012531}.
	
	\bibitem[Yuste(2010)]{yuste2010dendritic}
	Rafael Yuste.
	\newblock \emph{Dendritic spines}.
	\newblock MIT press, 2010.
	
	\bibitem[Zimmermann et~al.(2015)Zimmermann, Santos, Kovar, and
	Rock]{zimmermann2015actin}
	Dennis Zimmermann, Alicja Santos, David~R Kovar, and Ronald~S Rock.
	\newblock Actin age orchestrates myosin-5 and myosin-6 run lengths.
	\newblock \emph{Current Biology}, 25\penalty0 (15):\penalty0 2057--2062, 2015.
	
	\bibitem[Zimmermann(1996)]{zimmermann1996accumulation}
	Herbert Zimmermann.
	\newblock Accumulation of synaptic vesicle proteins and cytoskeletal
	specializations at the peripheral node of {R}anvier.
	\newblock \emph{Microscopy Research and Technique}, 34\penalty0 (5):\penalty0
	462--473, 1996.
	
\end{thebibliography}

\end{document}